\documentclass[onecolumn]{emulateapj}

\newcommand{\rg}{r_{\rm g}}

\newcommand{\Qvis}{Q_{\rm vis}^+}
\newcommand{\Qrad}{Q_{\rm rad}^-}
\newcommand{\Qadv}{Q_{\rm adv}^-}
\newcommand{\rtrap}{r_{\rm trap}}

\slugcomment{2006/5/6 ApJ accepted version!}

\shorttitle{K. Watarai}
\shortauthors{New Analytical Formula for Slim Disks}

\begin{document}
\title{New Analytical Formula for Supercritical Accretion Flows}

\author{Ken-ya Watarai \altaffilmark{1,2} }

\affil{1 Astronomical Institute, Osaka-Kyoiku University,
Asahigaoka, Kashiwara-city, Osaka, 582-8582, Japan}
\affil{2 Research Fellow of the Japan Society for the Promotion of Science}
\email{watarai@cc.osaka-kyoiku.ac.jp}

\begin{abstract}
We examine a new family of global analytic solutions
 for optically thick accretion disks, which includes the supercritical accretion regime. 
We found that the ratio of the advection cooling rate,
 $Q_{\rm adv}$, to the viscous heating rate, 
 $Q_{\rm vis}$, i.e., $f=Q_{\rm adv}/Q_{\rm vis}$,
 can be represented by an analytical form dependent on the radius and the mass accretion rate.  
The new analytic solutions can be characterized by the photon-trapping radius, $\rtrap$,
 inside which the accretion time is less than the photon diffusion
 time in the vertical direction; the nature of the solutions changes significantly as this radius is crossed. 
Inside the trapping radius, 
 $f$ approaches $f \propto r^0$,
 which corresponds to the advection-dominated limit ($f \sim 1$), 
 whereas outside the trapping radius, the radial dependence of $f$ changes to $f \propto r^{-2}$,
 which corresponds to the radiative-cooling-dominated limit. 
The analytical formula for $f$ derived here smoothly connects these two regimes. 
The set of new analytic solutions reproduces well the global disk
 structure obtained by numerical integration over a wide
 range of mass accretion rates, including the supercritical accretion regime.  
In particular, the effective temperature profiles for our new solutions 
 are in good agreement with those obtained from numerical solutions. 
Therefore, the new solutions will provide a useful tool not only for
 evaluating the observational properties of accretion flows,
 but also for investigating the mass evolution of black holes 
 in the presence of supercritical accretion flows.
\end{abstract}


\section{Introduction}

Recent X-ray observations have revealed the existence of bright X-ray sources 
 whose luminosities in some cases greatly exceed the Eddington luminosity of neutron stars,  
 where the Eddington luminosity is given by 
$L_{\rm E} \equiv 4 \pi c G M/\kappa_{\rm es} 
 = 1.25 \times 10^{38} (M/M_\odot) {\rm erg~s^{-1}},$
where $c$, $G$, $M$, and $\kappa_{\rm es}$ are the speed of light, the gravitational
constant, the mass of the central object, and the electron scattering opacity,
respectively. 
One of the best examples of such sources is the microquasar GRS1915+105 in our galaxy. 
This source emits mainly in the X-ray band,
 and with a luminosity of $L_{\rm X} \sim 10^{39} {\rm erg~s^{-1}}$,
 in its high/very high state, close to the Eddington luminosity (Belloni et al. 1997, 2000). 
Narrow line Seyfert 1 galaxies (NLS1s) are also candidates for 
 near or super Eddington sources, with typical luminosities of up to
 $10^{44-45} {\rm erg~s^{-1}}$, 
 despite their relatively small estimated black hole masses,
 $M \sim 10^6 M_\odot$ (Bian \& Zhao 2004; Collin \& Kawaguchi 2004).

One of the more attractive explanations for such large luminosities is a supercritical accretion onto a black hole. 
The history of supercritical accretion theory is older than one might expect;
 for example, it is discussed by Abramowicz et al. (1978),
 Jaroszy\'{n}ski et al. (1980), Paczynski \& Wiita (1980), and others.  
However, most of the old models are dynamically/thermally unstable (Begelman \& Meyer 1982). 
One of the causes of an unstable solution is
 the lack of an advective cooling term in the energy equation. 
An advective flow for supercritical accretion transports the thermal energy into the black hole,
 so that the flow finally cools the disk and maintains it in a stable state. 
The first stable solution for supercritical accretion flow was proposed by 
Abramowicz et al. (1988), which is the so -called ``slim disk'' model 
 (see in detail, Kato, Fukue, and Mineshige 1998, hereafter KFM98).
Such a solution is seen as an optically thick flow with an accretion rate
exceeding the critical rate, 
$\dot{M}_{\rm crit} \equiv L_{\rm E}/c^2 
 = 1.3 \times 10^{17} (M/M_\odot) {\rm g~s^{-1}}=2\times
 10^{-9}M_{\odot} {\rm yr}^{-1}$. 
Several authors have developed this model and applied it to luminous objects  
(Szuszkiewicz et al. 1996, Wang et al. 1999 for Active Galactic Nuclei (hereafter AGN);
 Mineshige et al. 2000 for NLS1s;
 Watarai et al. 2000 for Galactic black hole candidates, hereafter GBHCs)
 where good agreement is found between predicted spectral properties and recent X-ray observations. 
Watarai et al. (2001) explicitly showed that a slim disk model can explain
 the thermal component of the spectrum observed in ultraluminous X-ray sources (hereafter ULXs). 
Ebisawa et al. (2003) performed direct spectral fitting 
 of the slim disk model to observations of ULXs,
 and concluded that the slim disk model is the preferred model
 for the famous ULXs IC342 X-1. 

Simple models that can reproduce the exact numerical solutions are 
 helpful to test the theoretical models observationally.  
Self-similar solutions (hereafter SSS) for optically thin advection-dominated accretion flows
 (hereafter ADAF) were developed by Narayan and Yi (1994, 1995)
 and shown to be a good approximation to numerical solutions for a wide range of parameters
 (Narayan et al. 1997). 
Another class of self-similar ADAF solutions comprises the adiabatic
inflow-outflow solutions (Blandford \& Begelman 1999), which incorporate the
effects of a jet or outflow. 
SSS of the convection-dominated accretion flow (CDAF) model
 have also been investigated by several authors
 who have compared numerical results and SSS
 (Narayan et al. 2000; Quataert \& Gruzinov 2000),
 and discussed their spectral properties (Ball et al. 2001). 

Self-similar solutions for the slim disk model were derived by Watarai and Fukue (1999) and Wang and Zhou (1999). 
They explicitly derived effective temperature profiles of $T_{\rm
eff}\propto r^{-1/2}$, which agree well with those derived from the numerical results
 (Watarai et al. 2000). 
(In the case of the standard model by Shakura and Sunyaev (1973),
 the solution has already been expressed in analytical form,
 since the basic equations are not differential equations but algebraic equations.)
However, it is not yet known whether other quantities,
 such as radial velocity, rotational velocity, and sound speed
 are reasonably well described by the SSS. 
One problem is that previous SSS differ from
 the numerical results shown in Figure 2 of Watarai and Mineshige (2001),
 and so we would like to find more useful solutions. 
One reason for this problem is that
 unlike the self-similar case in optically thin ADAFs, 
 radiative energy loss plays an important role in optically thick solutions
 because the transition between the advection dominated regime
 and the radiative cooling dominated regime is very smooth. 
In contrast, the transition between an outer 
 standard-type disk and an inner ADAF should be clear (see, e.g., Honma 1996). 
If such a solution can be found, it will be useful and sufficiently powerful 
 to enable us to understand the flow properties,
 since it requires no numerical techniques.
In this paper, we report new self-similar solutions that are a good approximation
 to numerical results over a wide range in radius,
 and we discuss the properties of the solutions.

In the next section,
 we introduce the old SSS,
 following the same method as that adopted by Wang and Zhou (1999). 
In subsection \ref{subsec:global}, we derive the analytic form of $f$ from the energy equation,
 and demonstrate the new analytic solutions by modifying the old solutions. 
In section \ref{sec:compare}, we present profiles of physical quantities as a function
 of radius and compare the new solutions with numerical results. 
We then discuss the applications of the new solutions 
from the observational point of view in section \ref{sec:discussion}.
The final section summarizes our conclusions.

\section{Canonical Self-Similar Solutions and New Analytical Solutions}

In this section,
 we derive SSS based on the formulation of Wang and Zhou
(1999) (see also Spruit et al. 1987; Narayan \& Yi 1994). 
We assumed that the whole disk structure is optically thick. 
As in previous papers, 
 we introduced the ratio of the advective cooling rate, $\Qadv$
 to the viscous heating rate, $\Qvis$, $f$ ($=\Qadv/\Qvis$). 

In the optically thin ADAF,
 radiated photons can escape from the disk immediately, 
 so that $f$ keeps constant ($f \sim 1$) throughout the whole of the disk. 
However, the ratio $f$ should have a radial dependence
 in optically thick solutions. 
This is because $f$ gradually changes
 as the radius decreases due to the radiative diffusion process. 
In the next subsection, 
 we seek self-similar solutions without specifying
 the function of $f$ as a first step.
After that we will demonstrate the analytic form of $f$
 as a function of the radius and the accretion rate
 in subsection \ref{subsec:global}.

\subsection{Basic Equations and Canonical Self-Similar Solutions}

We solve the following hydrodynamic equations, (i.e. mass
conservation, momentum equation in the radial direction, angular
momentum conservation, hydrostatic balance), with the energy
equation. Here we ignore the energy loss or the mass loss effects via an outflow. 
\begin{eqnarray}
\label{mass}
    \dot{M}
     &=& -2\pi r v_{r} \Sigma, 
\\
\label{r-mom}
     v_r \frac{dv_r}{dr}+ \frac{1}{\Sigma}\frac{d\Pi}{dr}
     &=& r (\Omega^2-\Omega_{\rm K}^2) 
         - \frac{\Pi}{\Sigma}\frac{d \ln \Omega_K}{dr},  
\\
\label{ang-mom}
    \dot{M} (\ell-\ell_{\rm in}) 
     &=& -2 \pi r^2 T_{r \varphi}, 
\\
\label{hydro}
   (2N+3)
   \frac{\Pi}{\Sigma} 
     &=& H^2 \Omega_{\rm K}^2, 
\\
\label{energy}
   Q_{\rm vis}^+ 
     &=& Q_{\rm adv}^- + Q_{\rm rad}^-,  
\end{eqnarray}
where $\Sigma$, $\ell$, $\ell_{\rm in}$, and $H$ are the surface density
 , $\Sigma \equiv \int \rho_0 dz$,
 specific angular momentum ($\ell=r^2 \Omega$),
 angular momentum at inner edge of the disk, and the scale height.
And we adopt the Keplarian angular frequency in a Newtonian potential,
 $\Omega_{\rm K}=(GM/r^3)^{1/2}$. 
The $r$-$\varphi$ component of the height-integrated viscous stress tensor was expressed by $T_{r
\varphi}$, which is related to the total pressure by $T_{r \varphi}= -\alpha
\Pi$, where $\Pi$ is the height-integrated total pressure as defined by $\Pi
\equiv \int p dz$, 
and $\alpha$ is the viscosity parameter (Shakura and Sunyaev, 1973). 
We assumed a polytropic relationship, $p_0=K \rho^{1+1/N}_0$, in the vertical
 direction, where $N$ is the polytropic index. 

Viscosity generates thermal energy through the differential rotation of turbulently moving gas, 
 and the viscous heating rate is expressed by 
\begin{equation}
\label{qvis}
Q_{\rm vis}^{+} = - r T_{r \varphi} \frac{d\Omega}{dr} 
 \approx  -T_{r \varphi} \Gamma_{\Omega} \Omega,
\end{equation}
where $\Gamma_{\Omega}$ is the linear approximation form of angular velocity ($\Gamma_{\Omega}=- d \ln\Omega/d \ln r $).
Equation (\ref{qadv}) represents the ``net'' internal energy flux at
each radius; i.e., advective cooling. 
\begin{equation}
\label{qadv}
Q_{\rm adv}^{-} = - \frac{\dot{M}}{2\pi r} T_0 \frac{ds_0}{dr} 
 \approx \frac{\dot{M}}{2 \pi r^2} \frac{\Pi}{\Sigma} \xi.
\end{equation}
Variables, $T_0$ and $s_0$
 represent the temperature and the entropy in the equatorial plane,
 and $\xi$ is a dimensionless quantity given by
\begin{equation}
\xi = - \frac{1}{\Gamma_{3}-1}
 \left(\frac{d\ln{p_0}}{d\ln{r}} - \Gamma_3 \frac{d\ln{\rho}_0}{d\ln{r}} \right),
\end{equation}
where $\Gamma_3$ is the ratio of the generalized specific heats (e.g., Chandrasekhar 1967; Kato et al. 1998). 
Here $\xi$ is of order unity, and so we set $\xi = 1.5$ for the remainder of this paper. 

We assumed the disk to be radiation-pressure -dominated,
 with opacity due to electron scattering only; 
\begin{equation}
\label{qrad}
Q_{\rm rad}^- = \frac{8 a c T_{\rm 0}^4}{3 \kappa \rho H}
      \simeq \frac{8c \Pi}{\kappa_{\rm es} \Sigma H}. 
\end{equation}
where $\kappa_{\rm es}$ is the electron scattering opacity,
 $\kappa \simeq \kappa_{\rm es}$.

To derive the characteristic radial dependence of the above equations, 
we transformed the equations (\ref{r-mom}), ($\ref{ang-mom}$),
 (\ref{qvis}), (\ref{qadv}) as follows;
\begin{equation}
\label{qvis2}
Q_{\rm adv}^{-} = fQ_{\rm vis}^{+} 
\Longrightarrow 
 \frac{\Pi}{\Sigma}
= \frac{B \Gamma_\Omega \Omega_0^2}{\xi} f r^2 \Omega_{\rm K}^2,
\end{equation}
where $B=(1-\ell_{\rm in}/\ell)$ is the boundary term. 
We also assumed $\Omega=\Omega_0 \Omega_{\rm K}$,
 where $\Omega_0$ is a constant. 
The momentum equations in the radial and azimuthal direction are  
\begin{equation}
\label{r-mom2}
      \Gamma_{v} v_{r}^2 
    + \left(\Gamma_{\Pi} - \frac{3}{2} \right)\frac{\Pi}{\Sigma} 
    - r^2 (\Omega^2-\Omega_{\rm K}^2) = 0,  
\end{equation}
\begin{equation}
\label{ang-mom2}
   \dot{M} \Omega_0 \Omega_{\rm K} B = 2 \pi \alpha  \Pi,
\end{equation}
where $\Gamma_{v}=d \ln v_r/d \ln r $, $\Gamma_{\Pi}=d \ln\Pi/d \ln r $, and
 the coefficients were determined from the SSS. 
We assumed $\Omega_{0} \approx B \approx 1$ and $N=3$ for simplicity. 
Since the above equations can be solved analytically,
 we finally obtained the following solutions: 
\begin{eqnarray}
\label{ss-omega}
 \Omega &=& \Omega_0 \Omega_{\rm K}, \nonumber \\
   &\approx& 7.17 \times 10^{3} \left(\frac{\Omega_0}{1.0}\right) \left(\frac{m}{10}\right)^{-1} \left(\frac{r}{\rg}\right)^{-3/2} {\rm s^{-1}}, \\
\label{ss-vr}
 |v_r| &=& \left(\frac{\Gamma_\Omega \Omega_0 }{\xi }\right) f \alpha  v_{\rm K}, \nonumber \\
   & \approx & 3.17 \times 10^{9} f \left(\frac{\alpha}{0.1}\right) \left(\frac{r}{\rg}\right)^{-1/2} {\rm cm~s^{-1}}, \\
\label{ss-sigma}
 \Sigma &=& \left( \frac{\xi}{2\pi \Gamma_{\Omega} \Omega_0} \right)
          f^{-1} \alpha^{-1} \dot{M} r^{-1} v_{\rm K}^{-1}, \nonumber \\
        & \approx & 2.36 \times 10^{3} f^{-1} \left(\frac{\alpha}{0.1}\right)^{-1}  \left(\frac{\dot{m}}{100}\right) \left(\frac{r}{\rg}\right)^{-1/2} {\rm g~cm^{-2}}, \\
\label{ss-pi}
 \Pi &=& \frac{\Omega_0 B}{2\pi} \dot{M} \alpha^{-1} \Omega_{\rm K}, \nonumber \\
   &\approx& 1.58 \times 10^{24} \left(\frac{\alpha}{0.1}\right)^{-1} 
\left(\frac{\dot{m}}{100}\right) \left(\frac{r}{\rg}\right)^{-3/2} {\rm dyn~cm^{2}}, \\
\label{ss-H}
 H &=& \left[ (2N+3) \frac{B \Gamma_{\Omega} \Omega_0^2}{\xi} \right]^{1/2}  f^{1/2} r, \nonumber \\
   &\approx& 1.01 \times 10^7 f^{1/2} \left(\frac{m}{10}\right) \left(\frac{r}{\rg}\right) {\rm cm}. 
\end{eqnarray}
The coefficients $\Gamma_{\Omega}$, $\Gamma_{v}$, and $\Gamma_{\Pi}$
 can be determined self-consistently from the above equations (Spruit et al. 1987).   
In the limit $f \to 1$, all of the solutions are consistent with
 Watarai and Fukue (1999) and Wang and Zhou (1999). 
To summarize the normalized parameters, the mass accretion rate is 
$\dot{m}=\dot{M}/\dot{M}_{\rm crit} = \dot{M}/(L_{\rm E}/c^2)$, 
the Schwarzschild radius is $\rg = 2GM/c^2$, and the black-hole mass is $m =
M/M_{\odot}$, respectively.

We assumed that the equation (\ref{ss-pi}) is equal to the height-integrated
 radiation pressure, 
\begin{equation}
\Pi
= \Pi_{\rm rad}
= 2 I_{4} \frac{a}{3}T_0^4 H,
\end{equation}
where $a$ is the radiation constant and $I_4$(=128/315) is a numerical
constant (H\={o}shi 1977). 
The radiative flux at each radius is  
\begin{equation}
F = \frac{1}{2} \Qrad = \frac{16 \sigma T_0^4}{3\tau} =
\sigma T_{\rm eff}^4.
\end{equation}
Here $\Qrad$ is the radiative cooling rate in the optically thick
medium (diffusion approximation).
The optical depth $\tau$ was defined by $\tau = \kappa_{\rm es} \Sigma/2
$ because the opacity is assumed to be electron-scattering-dominated. 
Hence, we can estimate the effective temperature distribution; 
\begin{eqnarray}
\label{eq:Teff}
T_{\rm eff} &=& \left[\frac{8}{\kappa_{\rm es} I_4 a} \sqrt{\frac{B \Gamma_{\Omega}}{(2N+3) \xi}}\right]^{1/4} f^{1/8} r^{1/4}\Omega_{\rm K}^{1/2}, \nonumber \\
&\approx& 2.52 \times 10^7 f^{1/8} \left(\frac{m}{10}\right)^{-1/4} \left(\frac{r}{\rg}\right)^{-1/2} K.
\end{eqnarray}
This temperature profile is characteristic of an optically thick ADAF. 
The slim disk behavior is expressed well by equation (\ref{eq:Teff}).
The radial temperature profile is flatter than that of
the standard disk (changing from $T_{\rm eff} \propto r^{-3/4}$ for standard disks
 to $T_{\rm eff} \propto r^{-1/2}$ for slim disks).  
It is interesting to note that the effective temperature does not depend
on the accretion rate in this solution. 
However, the parameter $f$ should be a function of the mass accretion rate 
 because $f \approx 1$ corresponds to the high $\dot{m}$ limit in the disk. 
That is, the effect of $\dot{m}$ is implicitly included in $f$. 

\subsection{Global Solutions including a Radiative-Cooling-Dominated Regime}
\label{subsec:global}

According to the numerical results of Watarai and Mineshige (2001), 
 $f$ profiles depend strongly on $\dot{m}$ and radius. 
Strictly speaking, 
 the energy equation used in the previous subsection
 is inconsistent because the adopted energy equation,
 $Q_{\rm adv}^- = f Q_{\rm vis}^+$, does not explicitly include the radiative cooling term, $Q_{\rm rad}^-$. 
The full energy equation should be $Q_{\rm adv}^- = Q_{\rm vis}^+ - Q_{\rm rad}^-$, 
and so in the following analysis we start from this equation. 

First we included the effect of radiative cooling on $f$ using the following form; 
\begin{equation}
f = \frac{Q_{\rm adv}^-}{Q_{\rm vis}^+} = \frac{Q_{\rm adv}^-}{Q_{\rm adv}^-+Q_{\rm rad}^-}=\frac{1}{1+g}, 
\label{eq:f}
\end{equation}
where $g$ is the ratio of radiative cooling to advective cooling. 
The explicit form of $g$ is 
\begin{equation}
g \equiv \frac{Q_{\rm rad}^-}{Q_{\rm adv}^-} 
  = \frac{16 \pi c r^2}{\kappa_{\rm es} H \dot{M} \xi}
  = \frac{8}{\sqrt{(2N+3) \xi B \Gamma_{\Omega}}\Omega_0 } \dot{m}^{-1} f^{-1/2} \hat{r}. 
\label{eq:g}
\end{equation}
By inserting equation (\ref{eq:g}) into equation (\ref{eq:f}) and (\ref{ss-H}),  
 we obtained a quadratic equation in $f$; 
\begin{equation}
 f^2 - (D^2 x^2 +2)f + 1 = 0,
\label{eq:2ji}
\end{equation}
 where $D$ and $x$ are numerical coefficients
 ($D = 8/\sqrt{(2N+3) \xi B \Gamma_{\Omega}} \Omega_0$,
 for example, $D \approx 2.18$ for $N=3$, and $\Gamma_{\Omega}=1.5$), 
 and there is a dependence on the radius and accretion rate ($x= \hat{r}/\dot{m}$).  
This method is similar to an iteration procedure, 
 and so we refer to it as a ``semi-iterative method.'' 
 
It is clear from the discriminant of equation (\ref{eq:2ji}) that 
 although two real roots exist for equation (\ref{eq:2ji}),
 the range of $f$ is physically constrained to $0 \le f \le 1$.  
We can finally obtain an analytical solution for $f$; 
\begin{equation}
 f(\dot{m}, \hat{r}) = f(x) = 0.5 \left[D^2 x^2 +2 -D x
 \sqrt{D^2 x^2 +4} \right].
\label{eq:solf}
\end{equation}
In the limit $x \to 0$, $f(x)$ tends toward unity, which corresponds to a high accretion rate disk. 
At the other extreme, i.e. $x \to \infty$, $f(x) \to 0$,
 and this case corresponds to a low accretion-rate disk. 
Note that in the above equation, $f$ is a function of the radius
 and mass accretion rates. 
This result is completely different from
 previous self-similar ADAF solutions. 
\begin{figure}
  \begin{center}
   \epsscale{1.0}
    \plotone{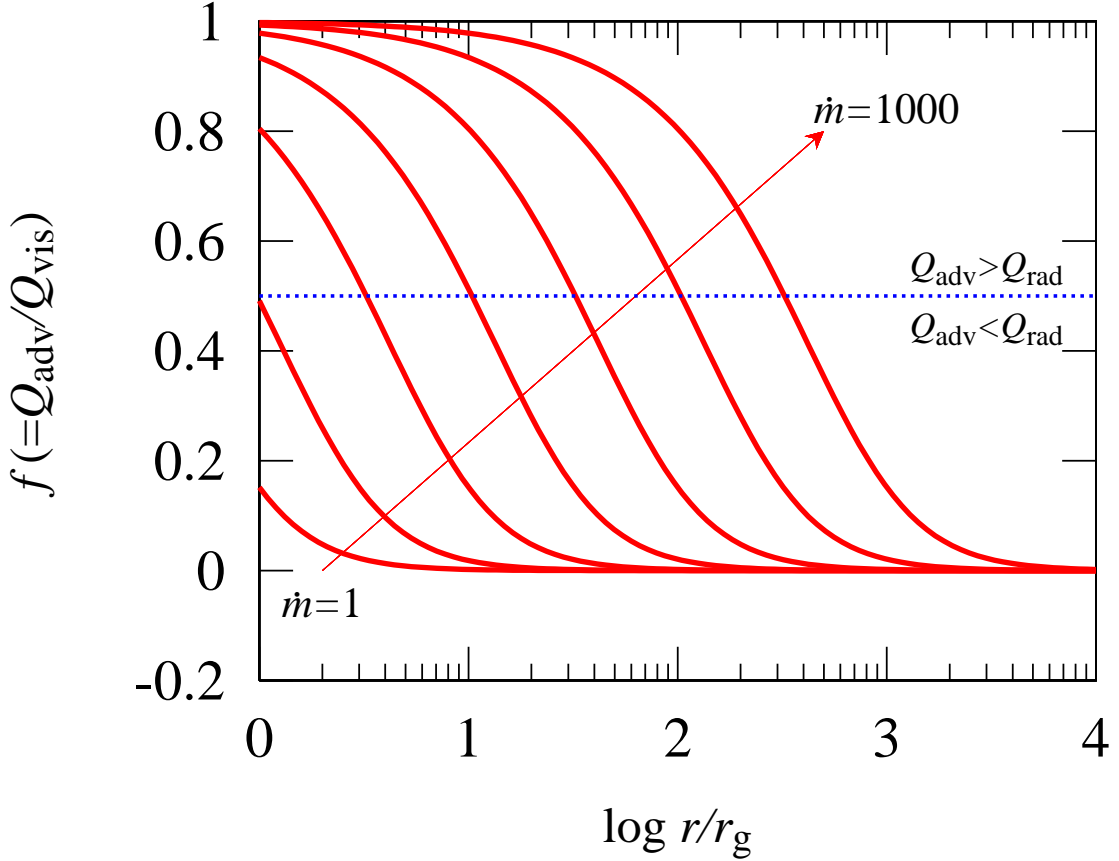}
  \end{center}
  \caption{The ratio of the advective cooling rate to the viscous heating rate.
 Solid lines represent the analytic expression of $f$ in equation (\ref{eq:solf}).  
By definition, the flow is advection-dominated above the dotted line ($f = 0.5$). 
Input mass accretion rates are $\dot{m}=10^0,~10^{0.5},~10^{1},~10^{1.5},~10^{2},~10^{2.5}~{\rm and} ~10^{3}$ 
 from bottom to top.}
\label{fig:f}
\end{figure}
Figure \ref{fig:f} represents the radial distribution of $f(\dot{m},\hat{r})$
 for various accretion rates. 
The value of $f$ goes to zero for smaller accretion rates and larger disk radii. 
For larger $\dot{m}$ and smaller $r$, the $f$ approaches to unity,
 i.e., becomes fully advection-dominated. 
These solutions are consistent with 
 radiation-pressure-dominated disks.
Note that the function $f$ smoothly connects the advection-dominated regime
 and radiation-dominated regimes. 
When we replace $f$ in equation (\ref{ss-vr}), (\ref{ss-sigma}), (\ref{ss-H})
 with $f(\dot{m},\hat{r})$, the physical quantities of the advective-cooling-dominated regime
 are integrated with those of the outer radiative-cooling-dominated regimes.

\subsection{Photon-Trapping Radius}

For high accretion rates, some fraction of the photons that are generated in
the disk cannot escape from the disk surface because they are trapped by
 the radial flow, so that the photons fall into the black hole along with the gas.  
It is natural that the efficiency of the conversion of gravitational
potential energy to radiation energy decreases in this situation  
 (Begelman 1978; Begelman \& Meyer 1982). 

In the 1D case, we can roughly estimate the radius of photon trapping, 
$\rtrap$, from a timescale argument. 
In order for photon trapping to occur, the accretion time must be shorter
 than the photon diffusion time, $t_{\rm acc} \lesssim t_{\rm dif}$, 
 where $t_{\rm acc}$ is the accretion time ($\sim r/v_r$),
 $t_{\rm dif}$ the diffusion time in the vertical direction ($\sim
 H/(c/\tau)$), and $\tau$ is an optical depth.  
We derived the trapping radius, $r_{\rm trap}$, as follows;  
\begin{equation}
\rtrap \approx \frac{H}{c}\tau v_r 
= \sqrt{\frac{B \Gamma_\Omega \Omega_0^2}{2 \xi}} 
  \frac{\kappa_{\rm es}}{4 \pi c} f^{1/2} \dot{M},  
\end{equation}
so that the normalized trapping radius is, 
\begin{equation}
\hat{r}_{\rm trap} = \frac{\sqrt{3}}{2} f(\hat{r}_{\rm trap}, \dot{m})^{1/2} \dot{m}.
\label{trapr}
\end{equation} 
The previous equation is a biquadratic equation of $\hat{r}_{\rm trap}$, and so we can solve it analytically,  
\begin{eqnarray}
 \hat{r}_{\rm trap} &=& \sqrt{-\frac{3}{3 D^2-4}+\frac{1}{2}\sqrt{\frac{144}{(3D^2-4)^2}+\frac{9}{3D^2-4}}} \dot{m}, \\
& \approx & 51 \left(\frac{\dot{m}}{100}\right) ~~({\rm for}~D = 2.18). 
\end{eqnarray}
Our analytic solutions roughly reproduce the numerical results  
 not only for the inner advection-dominated regime ($f \propto r^0$ for $r \lesssim r_{\rm trap}$)
 but also in the outer radiation-dominated regime ($f \propto r^{-2}$ for $r \gtrsim r_{\rm trap}$)
 except in the inner/outer boundary regions (see figure \ref{fig:energy}). 

Note that the trapping radius depends only on the accretion rate 
 and not on the other disk parameters. 
That is, this radius is the transition radius between the radiative-
cooling-dominated disk (standard disk) and the slim disk. 
The old SSS described in the previous subsection
 (i.e., the $f$=const. solutions) are valid
 $only~inside~this~radius$. 
However, the derived SSS may not be reliable
 in the disk's inner region due to the condition used for its inner boundary. 
It is therefore necessary to be careful about the useful range of the solutions. 

\subsection{Solutions in a Radiative-Cooling-Dominated Regime ($r>r_{\rm trap}$)}

Outside the trapping radius,
 we simply adopted the standard solution of Shakura and Sunyaev (1973). 
Our solutions are basically the same as the standard ones
 so that the viscous energy input balances the radiative cooling, 
 $\Qvis \simeq \Qrad$, and the disk is radiation-pressure-dominated. 
\begin{equation}
\label{}
\Qvis \simeq \Qrad 
\Longrightarrow 
 \frac{B \Gamma_\Omega}{2 \pi} \dot{M}\Omega^2 
\simeq \frac{8c \Pi}{\kappa_{\rm es} \Sigma H}. 
\end{equation}
From the hydrostatic balance and angular momentum equations, 
 we can rewrite the solutions as follows: 
\begin{eqnarray}
 v_r &=& - \frac{A_1^2 A_3 \Omega_0^3}{A_2^2} \alpha r^{-1} \Omega_{\rm K} \dot{M}^2  \nonumber \\
     &\approx& -5.96 \times 10^{12} 
\left(\frac{\alpha}{0.1}\right)
\left(\frac{\dot{m}}{100}\right)^2 
\left(\frac{r}{\rg}\right)^{-5/2}  {\rm cm~s^{-1}}, \\
 \Sigma &=& \frac{B}{2\pi} \left(\frac{A_2^2}{A_1^2 A_3 \Omega_0^3}\right)\alpha^{-1} \Omega_{\rm K}^{-1} \dot{M}^{-1} \nonumber \\
   &\approx& 1.26  
\left(\frac{\alpha}{0.1}\right)^{-1}
\left(\frac{\dot{m}}{100}\right)^{-1} 
\left(\frac{r}{\rg}\right)^{3/2}  {\rm g~cm^{-2}}, \\
 H &=& \frac{A_1 A_3}{A_2} \Omega_0^{2} \dot{M} \nonumber \\
   &\approx& 4.43 \times 10^{8} \left(\frac{\dot{m}}{100}\right) 
\left(\frac{m}{10}\right) {\rm cm}, \\
 T_{\rm eff}
 &\approx& 4.79 \times 10^{7} \left(\frac{m}{10}\right)^{-1/4} 
\left(\frac{\dot{m}}{100}\right)^{1/4} 
 \left(\frac{r}{\rg}\right)^{-3/4} {\rm K}, 
\end{eqnarray}
where $A_1=B \Gamma_\Omega/(2\pi), A_2=8c/\kappa_{\rm es},
A_3=(2N+3)$=9 (for $N$=3) are numerical constants.   
We set $\Omega_0 \approx B \approx 1,~\Gamma_{\Omega} \approx 1.5$ for simplicity. 
These solutions are all nearly the same as the Shakura and Sunyaev
type solutions for the radiation-pressure dominated regime. 
Note again that these solutions can be adopted at
 $r > \rtrap$. 

Here we summarize the radial dependence of physical quantities for each
regime (see table \ref{tab:sum}). 
In the following discussion, we will neglect the solutions for the gas-pressure dominated regime
 because such solutions have already been published in many papers.

\begin{table}
  \caption{Summary of the radial dependence of the physical quantities}
  \label{tab:sum}
  \begin{center}
    \begin{tabular}{lllll}
     Regime & $v_r$ & $ \Sigma$ & $T_{\rm eff}$ & $H$ \\
    \hline
      I   \small{($Q_{\rm vis}^+ \simeq Q_{\rm adv}^-, p \simeq p_{\rm rad}$)}   & $\propto r^{-1/2}$   & $\propto r^{-1/2}$ 
          & $\propto r^{-1/2}$ & $\propto r^{1}$ \\
      II  \small{($Q_{\rm vis}^+ \simeq Q_{\rm rad}^-, p \simeq p_{\rm rad}$)}   & $\propto r^{-5/2}$   & $\propto r^{+3/2}$ 
          & $\propto r^{-3/4}$ & $\propto r^{0}$ \\
      III \small{($Q_{\rm vis}^+ \simeq Q_{\rm rad}^-, p \simeq p_{\rm gas}$)}  & $\propto r^{-2/5}$   & $\propto r^{-3/5}$ 
          & $\propto r^{-3/4}$ & $\propto r^{21/20}$ \\
    \hline
    \end{tabular}
  \end{center}
\end{table}

\section{Comparison with Numerical Solutions} 
\label{sec:compare}

\subsection{For Stellar Mass Black Holes}

\begin{figure}
  \begin{center}
   \epsscale{1.0}
    \plotone{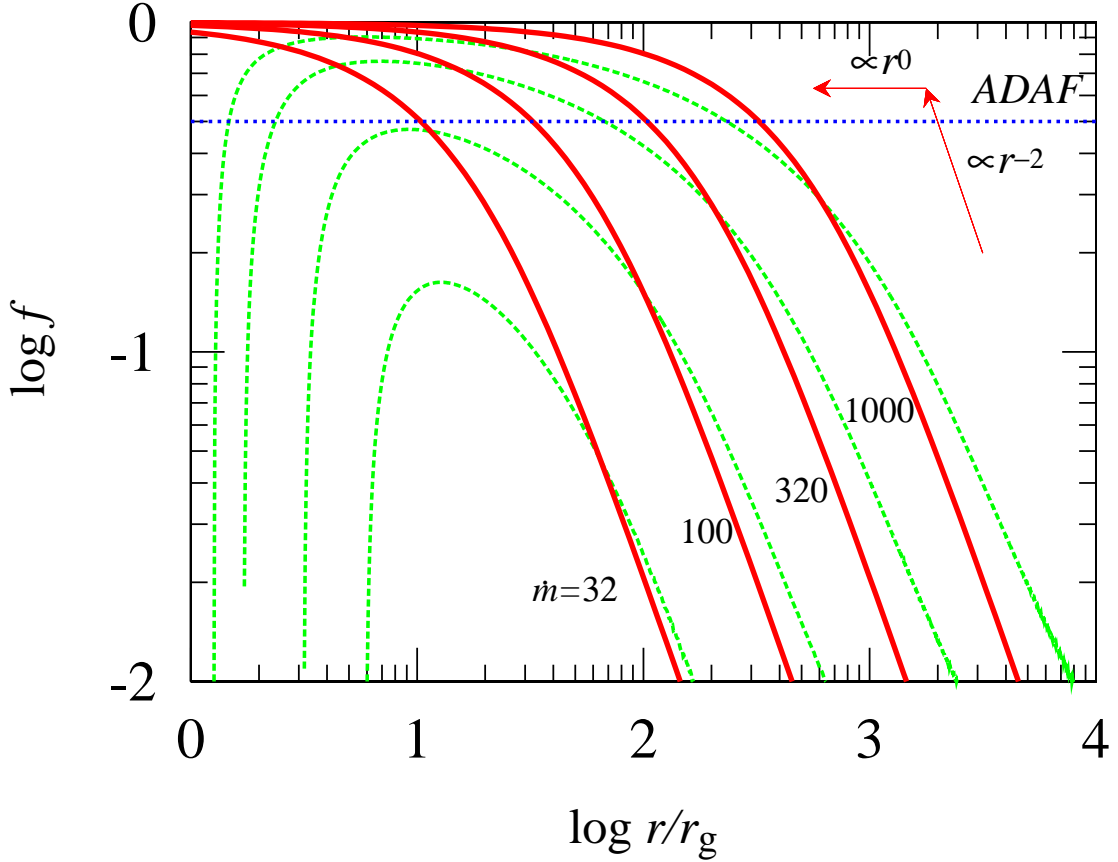}
  \end{center}
  \caption{Energy distributions of high $\dot{m}$ flows.
 Solid lines represent the analytical form of $f$ in equation (\ref{eq:solf}),
 and dashed lines show the numerical calculations. 
 The black-hole mass is fixed at $m=10$. 
By definition, the flow is advection-dominated above the dotted line ($f = 0.5$). }
\label{fig:energy}
\end{figure}

Figure \ref{fig:energy} shows a comparison
 of our new analytic solutions with numerical results. 
Numerical data are taken from the results of Watarai et al. (2000)
 and Watarai and Mineshige (2001). 
As is clear in figure \ref{fig:energy},
 equation (\ref{eq:solf}) reproduces  
 the $f$ profiles of the numerical solutions relatively well
 except for the inner/outer boundary region. 
The trapping radius derived from equation (\ref{eq:solf}) 
 tends to be overestimated by a factor of two to three due to inner boundary conditions; 
however, using the new solutions, $r_{\rm trap}$ is roughly consistent
 with numerically obtained values for higher mass accretion rates. 

\begin{figure}
  \begin{center}
     \epsscale{0.4}
    \plotone{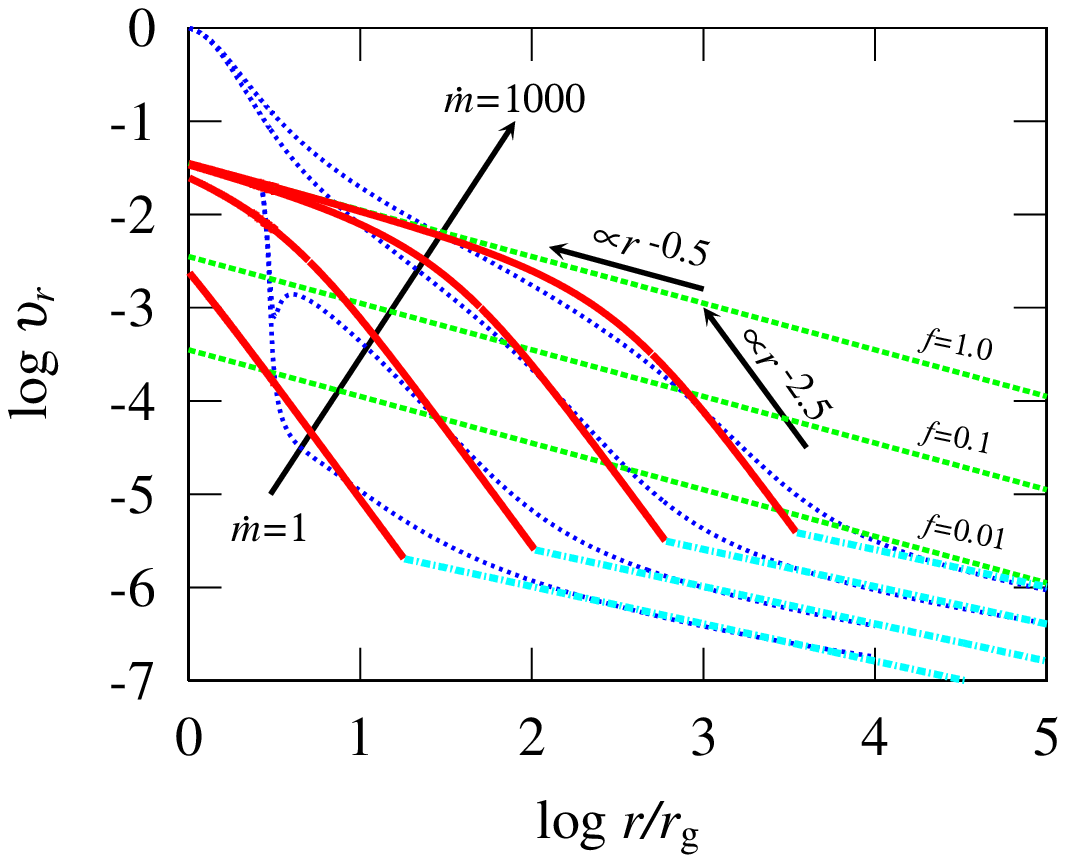}
    \plotone{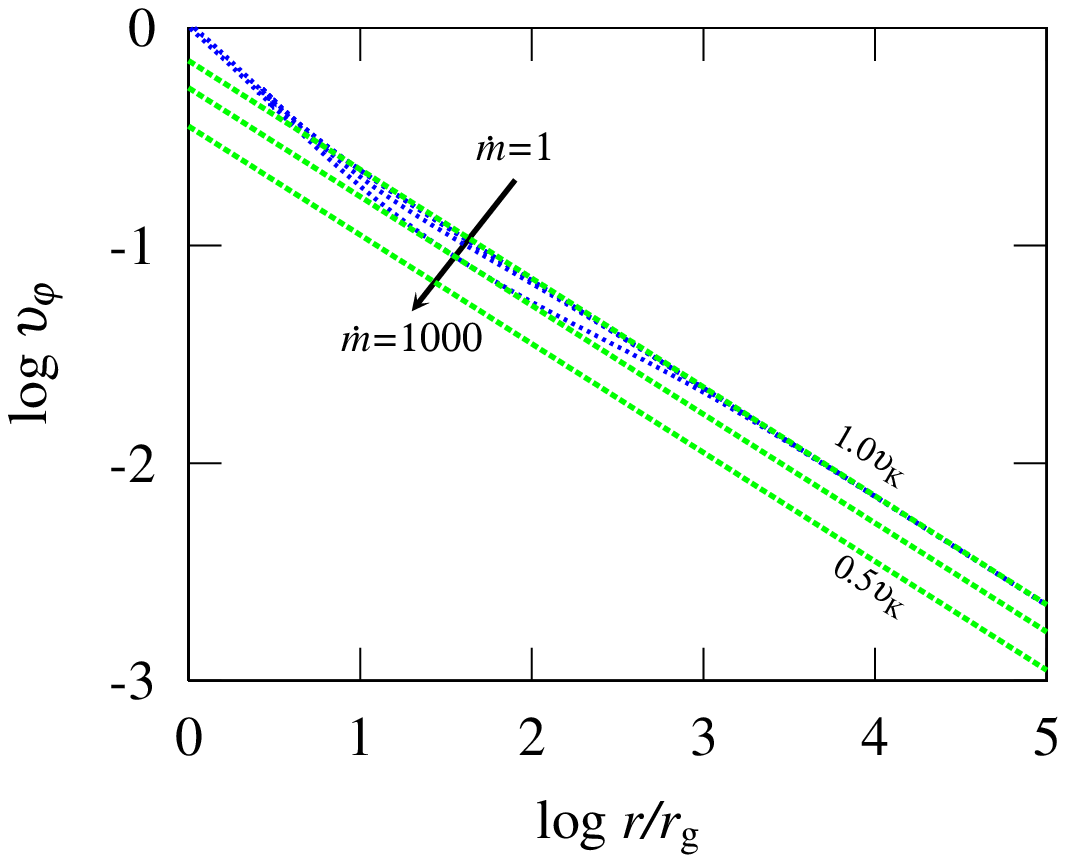}
    \plotone{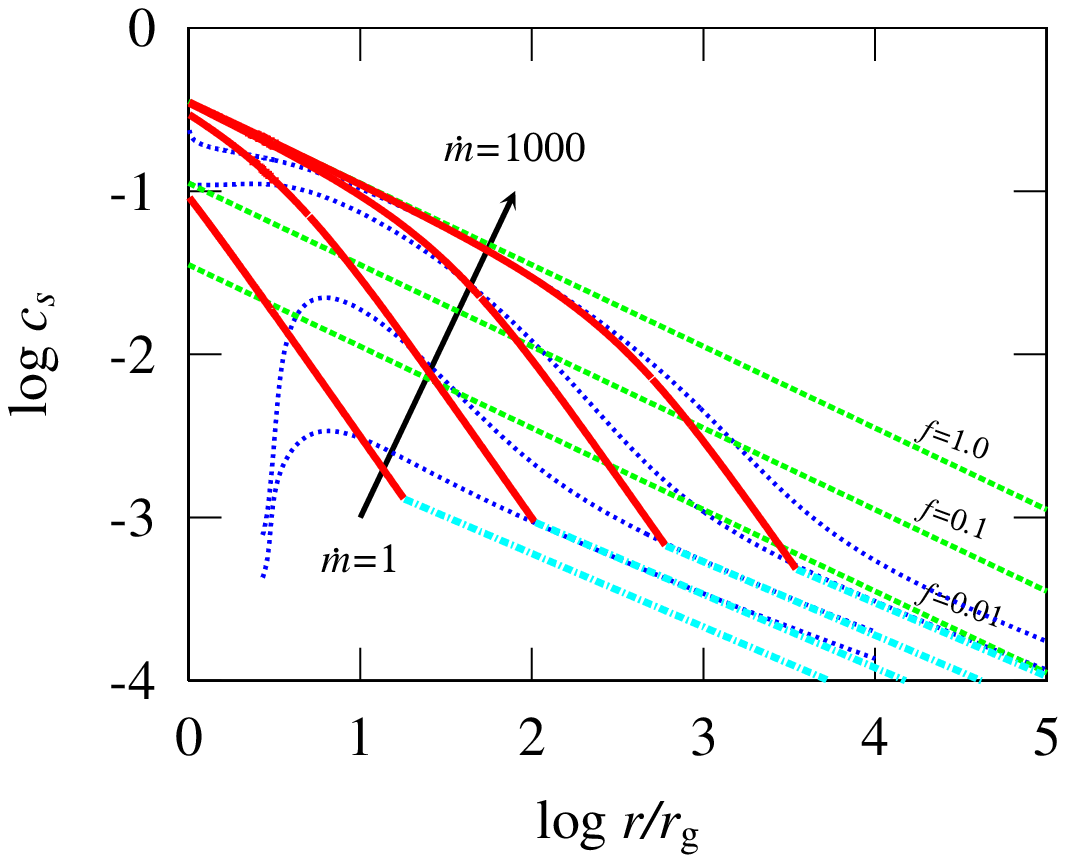}
    \plotone{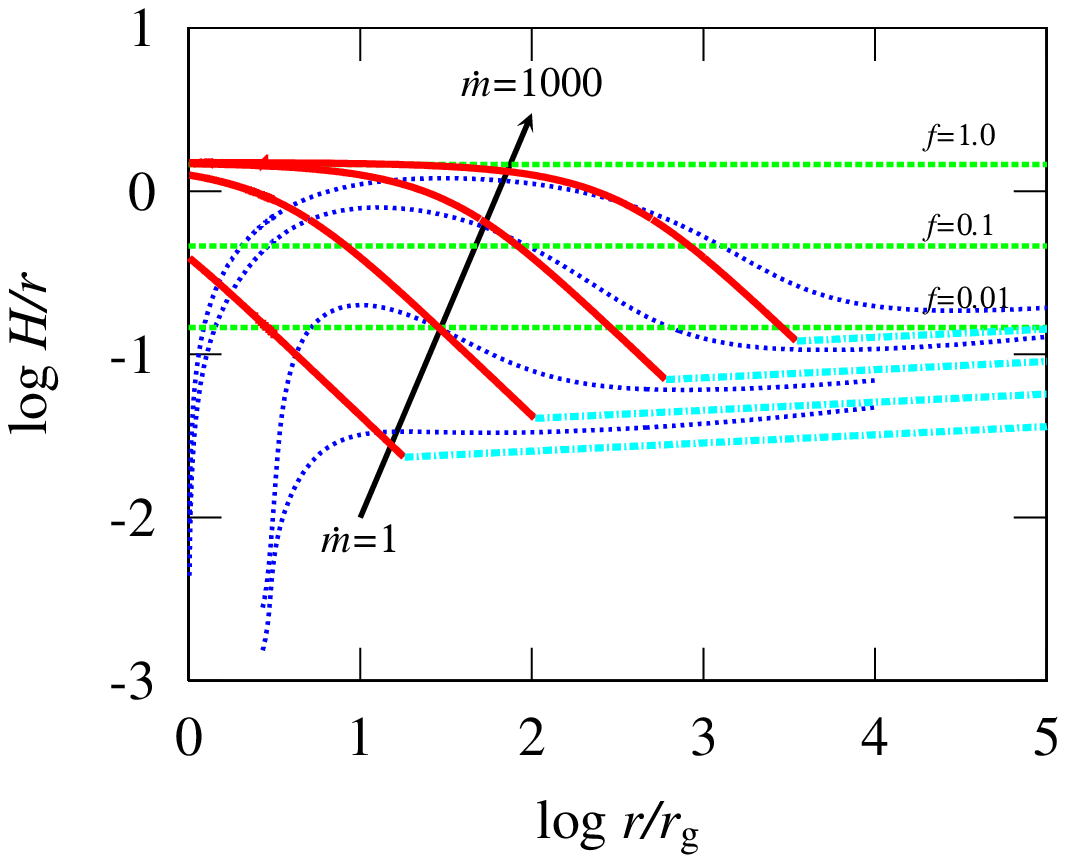}
    \plotone{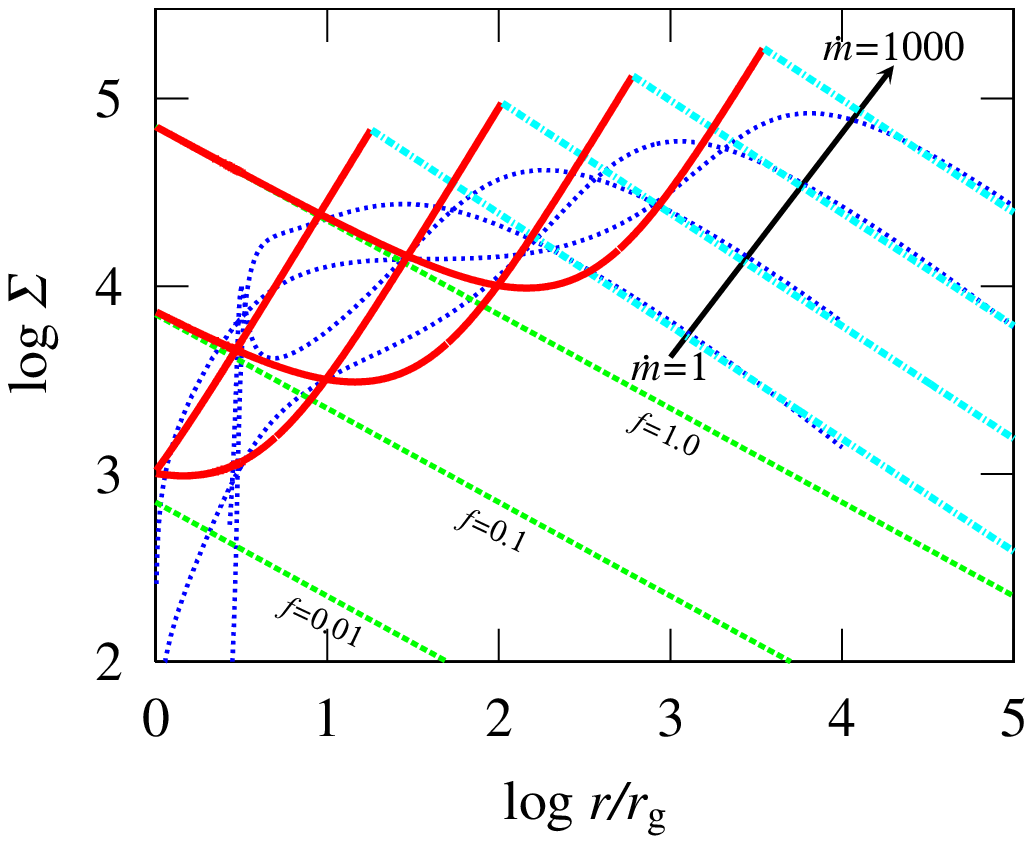}
    \plotone{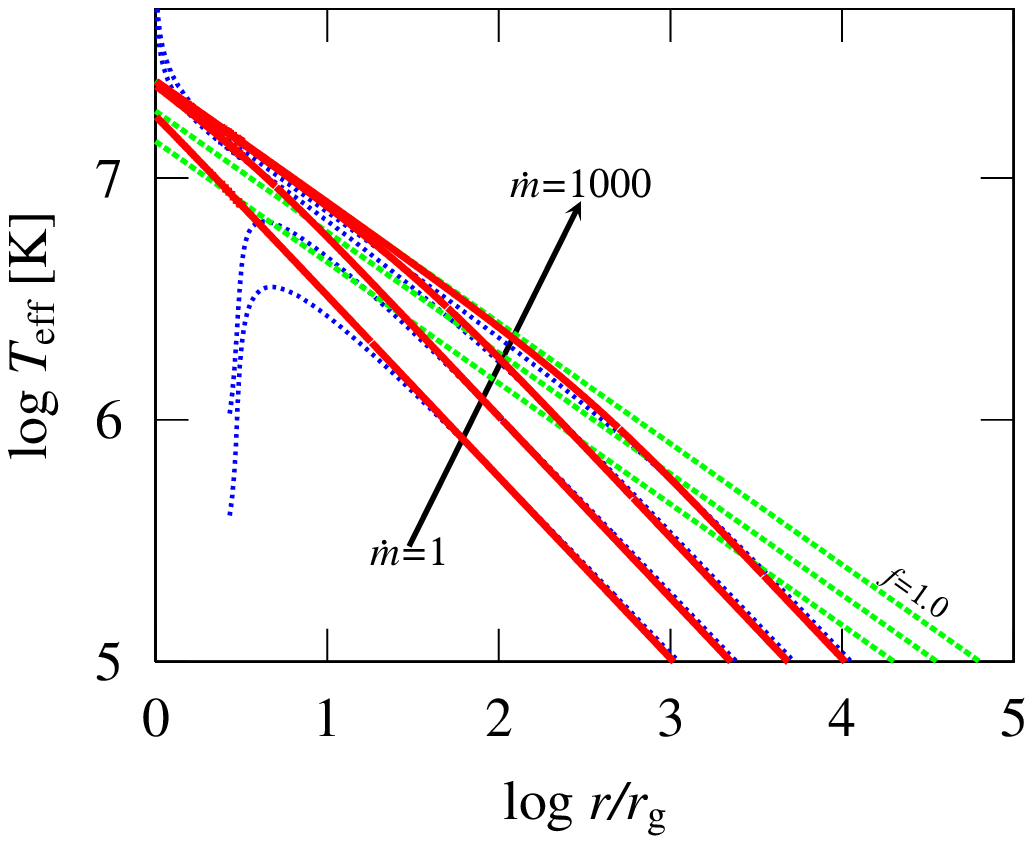}
  \end{center}
  \caption
{Radial velocity (upper left), azimuthal velocity (upper right),
 sound speed (middle left), scale height (middle right),
 surface density (bottom left), and effective temperature (bottom right) profiles. 
The thick solid lines represent
 the effect of variation of accretion rates $\dot{m}$=1, 10, 100, 1000 for
 our new analytic solutions, and the dashed lines are numerical solutions. 
The dotted lines show the variation of the parameter $f$=0.01, 0.1,
 1.0 for previous SSSs. 
The dot-dashed lines represent the gas-pressure-dominant regime at each accretion rate. 
Velocities are normalized by light speed. 
}
  \label{fig:profile1}
\end{figure}

Note that the numerical solutions do not always correspond to an advection-dominated flow 
even if supercritical accretion takes place ($\dot{m} \gtrsim 16$, 
see also figure 2 in Watarai \& Mineshige 2001). 
This shows that radiative cooling still works in the supercritical regime. 
Although this was demonstrated by Abramowicz et al. (1988), many authors
 do not recognize the fact that  
{\it the slim disk is not always advection-dominated; i.e.
 $Q_{\rm rad}^-$ is not always negligible.}
Therefore, when considering SSS in the regime of supercritical accretion, 
the radiative cooling term should be included. 
Figure \ref{fig:profile1} shows the various physical quantities obtained from the 
 new analytical solutions and numerical solutions
 for different accretion rates for a stellar mass black hole. 
The new solutions are indicated with thick solid lines,
 the numerical solutions with dashed lines,
 and the old SSS which correspond to $f=const.$ are indicated by dotted lines for comparison. 
All of the analytic solutions are smoothly connected
 from the advection-cooling-dominated regime to the radiative-cooling-dominated regime.

First, note that the velocity in the azimuthal direction roughly agrees with
the new analytic solutions (top right panel in figure \ref{fig:profile1}). 
The deviation of $v_\varphi$ between the analytic and numerical solutions
 is a factor of a few ($f \sim$ 0.1 at most). 
The important point is that the differences from the numerical solution are
 a factor of 2--3, and the radial dependence of $v_\varphi$ does not change to
$\propto r^{-1/2}$. 
This fact supports our initial assumptions that $\Omega \approx \Omega_{\rm K}$
 (i.e., $\Omega_0 \approx 1$) and $\Gamma_{\Omega} \approx 1.5$
 are valid over a large range of parameters. 

The radial- and sound- velocity, scale height,
 and effective temperature profiles obtained with the new analytic solutions
 are relatively consistent with those of the numerical solutions over a wide range of parameters. 
In particular, the effective temperature profiles
 are in excellent agreement with the numerical results
 for the whole disk structure for higher mass accretion rates ($\dot{m} \gtrsim 10$).  
This is because the effective temperature profile
 has a weak dependence on the inner-boundary-related quantities, i.e., 
$T_{\rm eff} \propto (B \Gamma_{\Omega}/\xi)^{1/8}$
 (see equation \ref{eq:Teff}). 

Discrepancies between the analytic solutions and numerical solutions
 appear at the disk's outer region in all quantities. 
This is because the outer region is in a gas-pressure-dominated state. 
Our analytical solutions assume a radiation pressure dominated state. 
Gas-pressure-dominated solutions have already been presented by 
 Shakura and Sunyaev (1973) (see also Kato et al. 1998), 
 and the boundary radius between the radiation-pressure dominated region and 
  gas-pressure dominated region is located at
\begin{eqnarray}
\label{rab}
\hat{r}_{\rm ab} & \simeq & 18 (\alpha m)^{2/21} \dot{m}^{16/21} \\
                 &\approx & 601 \left(\frac{\alpha}{0.1}\right)^{2/21} 
                          \left(\frac{m}{10}\right)^{2/21} 
                          \left(\frac{\dot{m}}{100}\right)^{16/21}. 
\end{eqnarray}
That is, if $r > r_{\rm ab}$, the disk is in the gas-pressure dominated state. 
Our new solutions are not useful in this regime.

Numerical solutions for $v_r$ have steeper profiles around the inner
boundary; however, the profile changes across the trapping radius.  
The standard radiation-pressure-dominated solution has a steep radial
profile, $\propto r^{-5/2}$, and so this profile agrees with the
numerical solutions. 
Strictly speaking, 
the numerical solutions are expressed by the combination of the solutions 
$\propto r^{-1/2}$ and $\propto r^{-5/2}$. 

The analytic solutions in the sound speed reproduce the solutions reasonably well. 
Here $c_s$ is derived from the relationship $c_{\rm s}=(\Pi / \Sigma)^{1/2}$. 
These results indicate that the thermodynamic quantities are expressed well
 by analytical solutions. 
These results can be explained by the fact that flow dynamics are 
influenced by radiative energy loss. 

The scale-height of the disk changes with radius, 
and our analytic solutions reproduce well the trend in disk structure.
The disk geometry is moderately thick (slim) inside $\rtrap$;
 outside $\rtrap$, in contrast, the disk is geometrically thin. 
As for the surface density profiles (bottom left panel), 
 the analytical solutions only reproduce the trend seen in the numerical solutions, 
 but they differ slightly in normalization. 
The effects of the inner/outer boundary
 are significant for the surface density profile. 

\begin{figure}
  \begin{center}
   \epsscale{1.0}
    \plotone{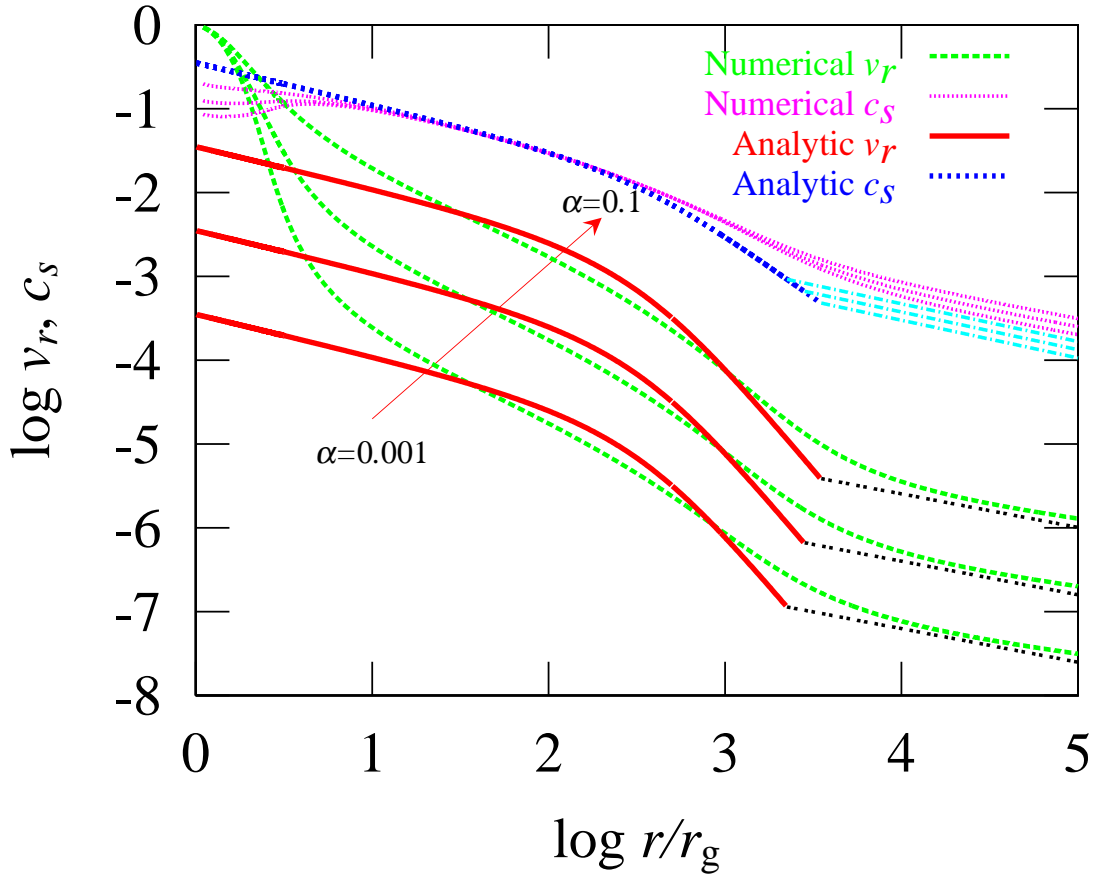}
  \end{center}
  \caption{Radial and sound velocity for different values of the viscous parameter $\alpha=0.001,~0.01,~0.1$ from bottom to top.
Mass accretion rate is set to be $\dot{m}=1000$. 
 Thick solid and thick dotted lines indicate
 the new analytic solutions for radial velocity, $v_{r}$, and sound velocity $c_{s}$, respectively. 
Dashed and small-dashed lines are numerical solutions. 
The dot-dashed lines around larger radii represent the gas-pressure-dominant regime. }
\label{fig:vr-alpha}
\end{figure}

As established in the classical model by Shakura and Sunyaev (1973), 
 the $\alpha$ dependence of physical quantities related to the disk dynamics, 
 density, surface density, and radial velocity is strong ($\rho, \Sigma \propto \alpha^{-1}$, $v_r \propto \alpha$),
 whereas the quantities related to the temperature
 are not sensitive to the viscosity (dynamics),
 i.e., $T_{\rm eff},~v_{\varphi}$ show no $\alpha$ dependence. 
($T_{\rm c}$, $c_{\rm s}$, and $H$ weak have a weak $\alpha$ dependence.) 
We also show the $\alpha$ dependence of the radial and sound velocities
 in figure \ref{fig:vr-alpha}. 
As mentioned above, 
 the $v_r$ are significantly affected by the value of $\alpha$,
 and the discrepancy between analytic and numerical solutions does not disappear, 
 however, the sound velocities from the analytic solutions agree well
 with those obtained from numerical calculations
 even if $\alpha$ varies by three orders of magnitude. 

Note that these analytic solutions are not exact solutions
 , but represent the characteristic properties
 of the equations.
However, the analytical solutions seem to be a good approximation 
 for the temperature and sound velocity profiles in the disk's inner region,
 which is dominated by advective cooling. 
Therefore, the analytic solution is a useful tool
 for estimating the bolometric luminosity in supercritical accretion disks. 

\subsection{For Supermassive Black Holes}

\begin{figure}
  \begin{center}
     \epsscale{0.45}
    \plotone{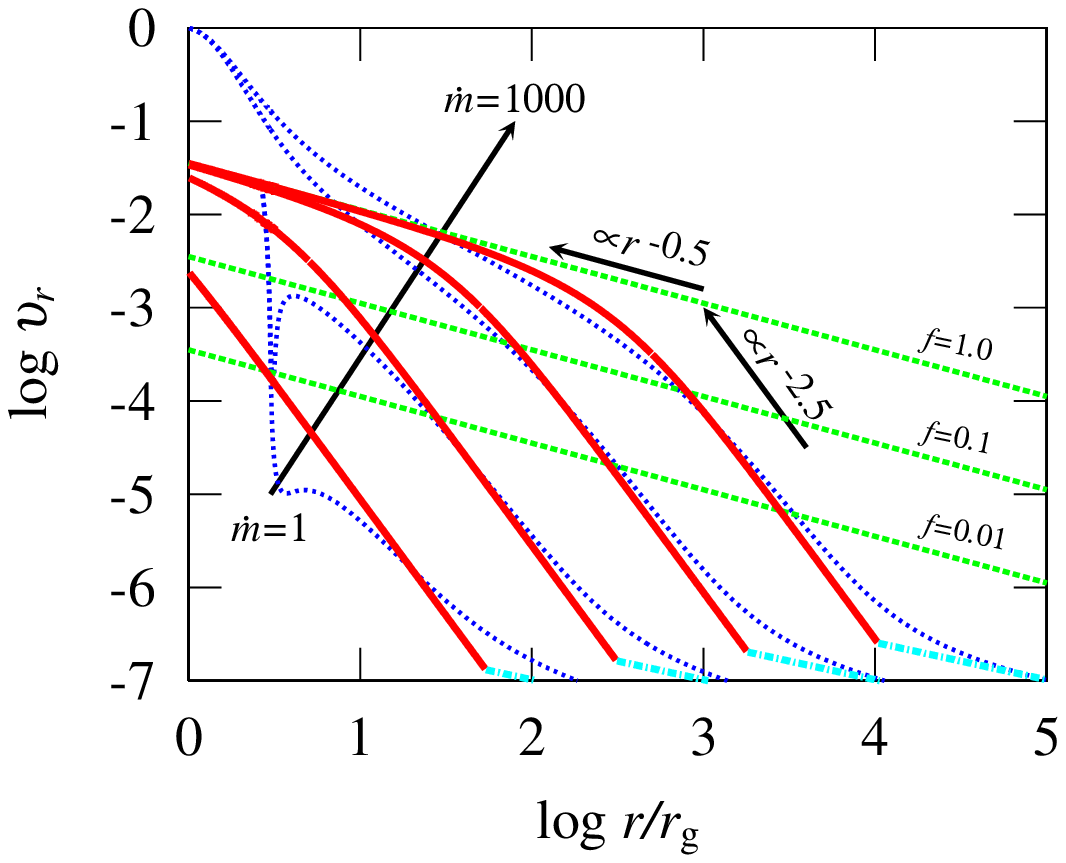}
    \plotone{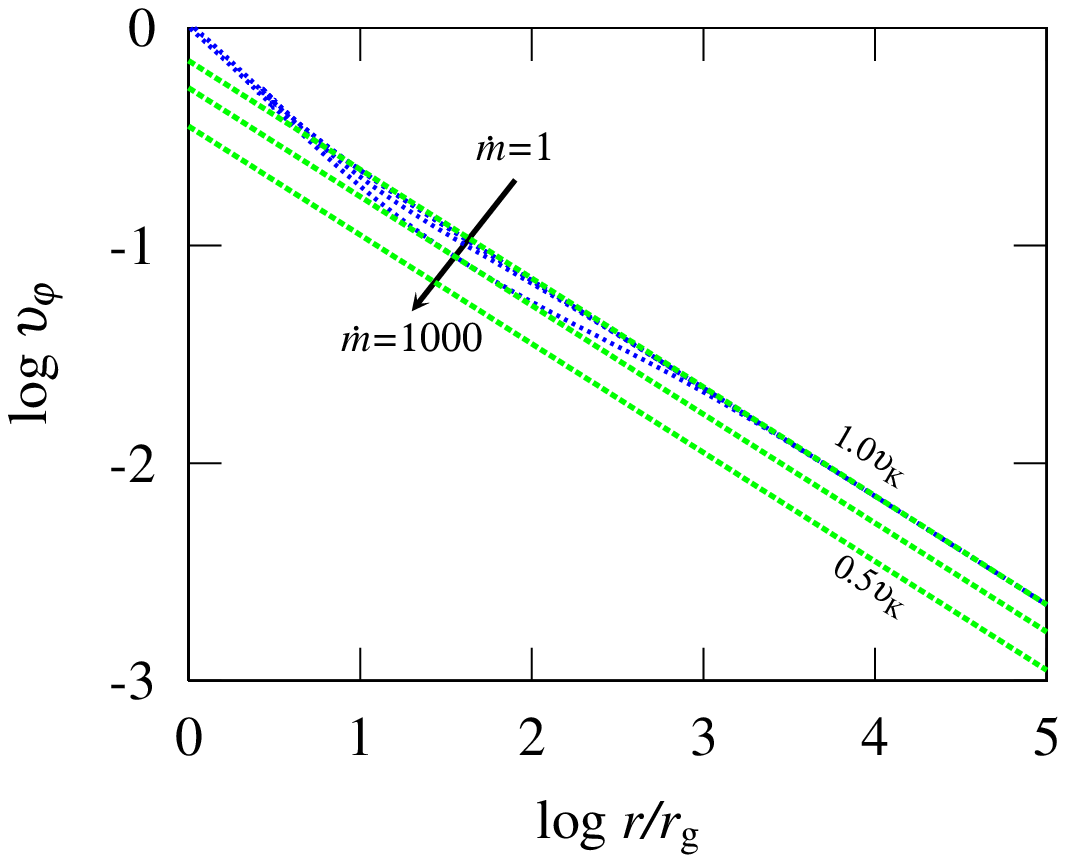}
    \plotone{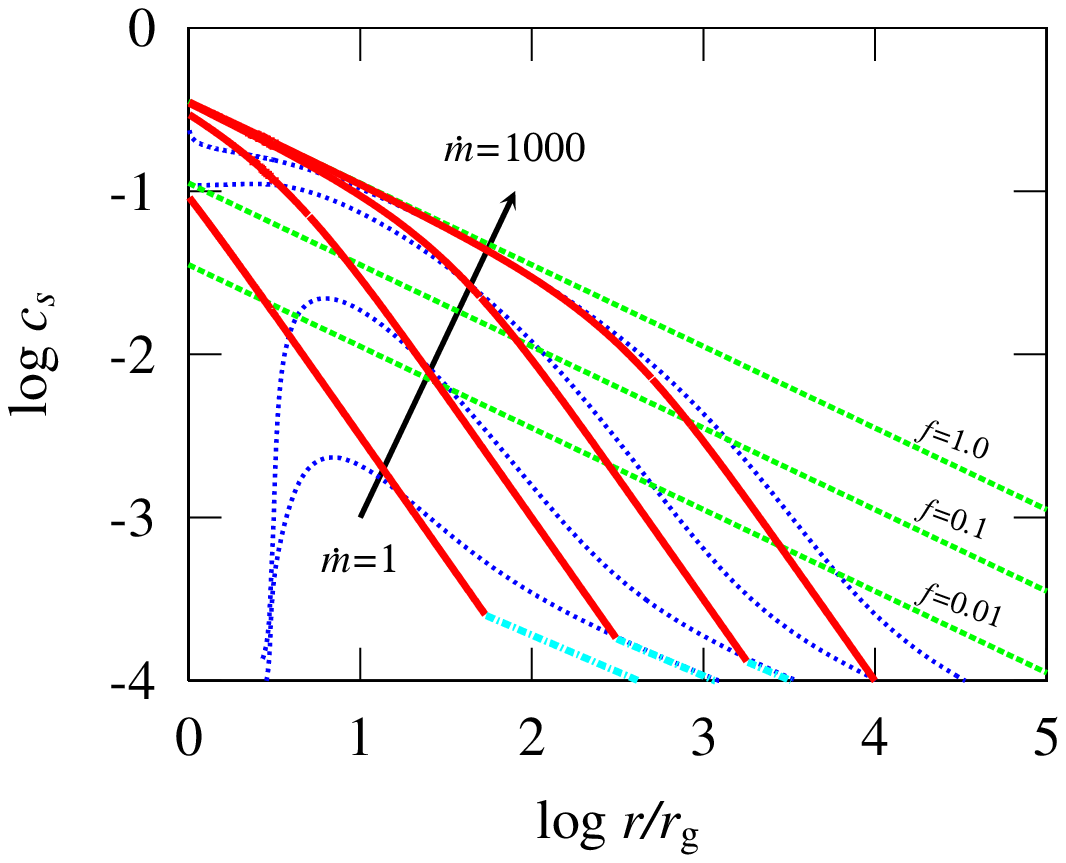} 
    \plotone{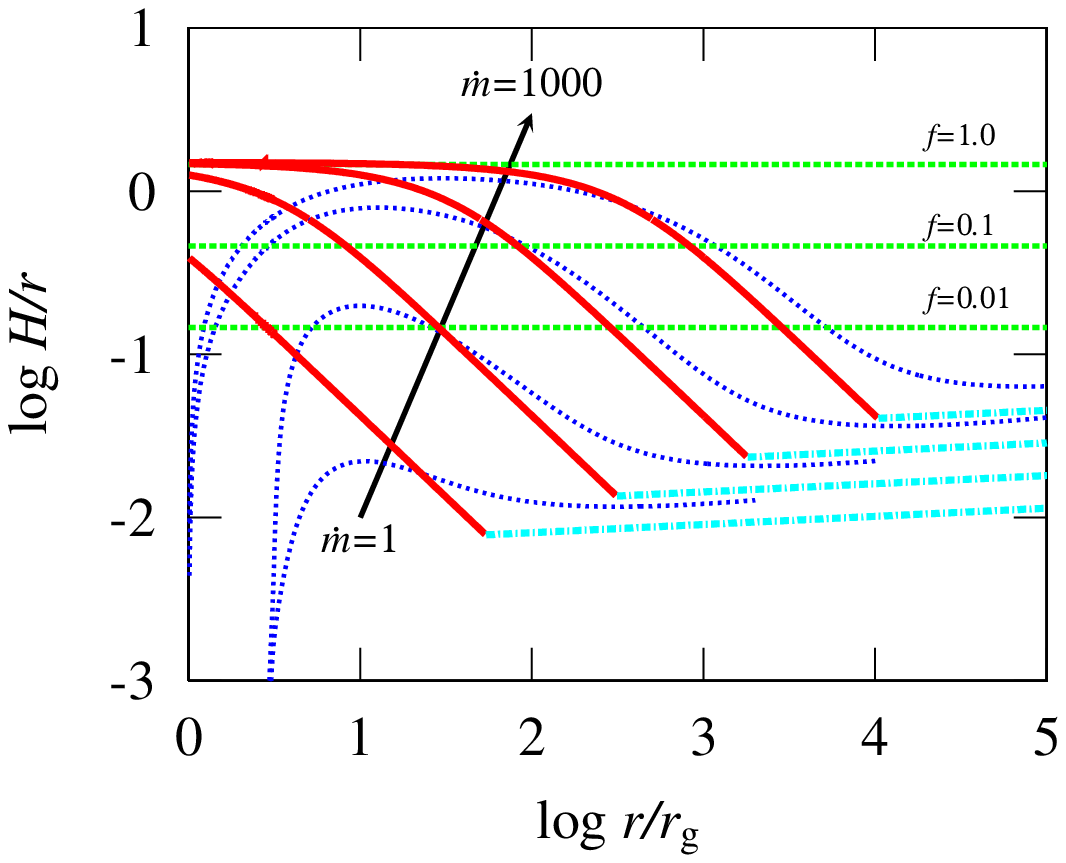}
    \plotone{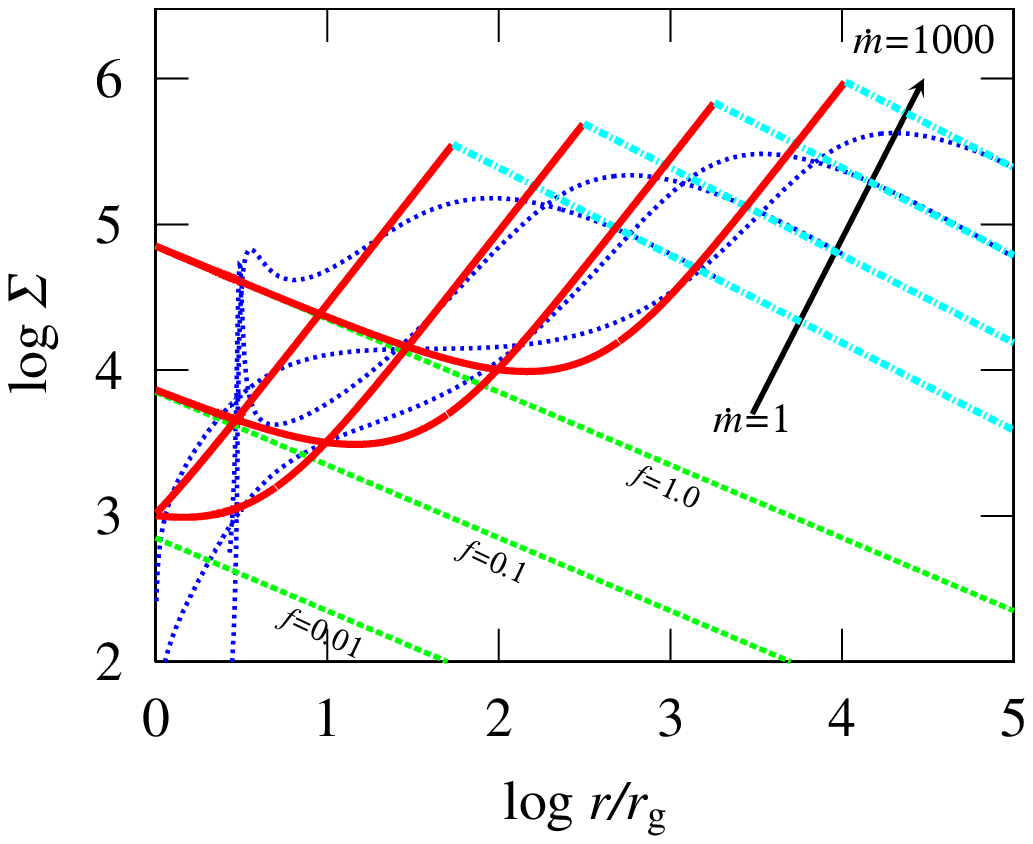}
    \plotone{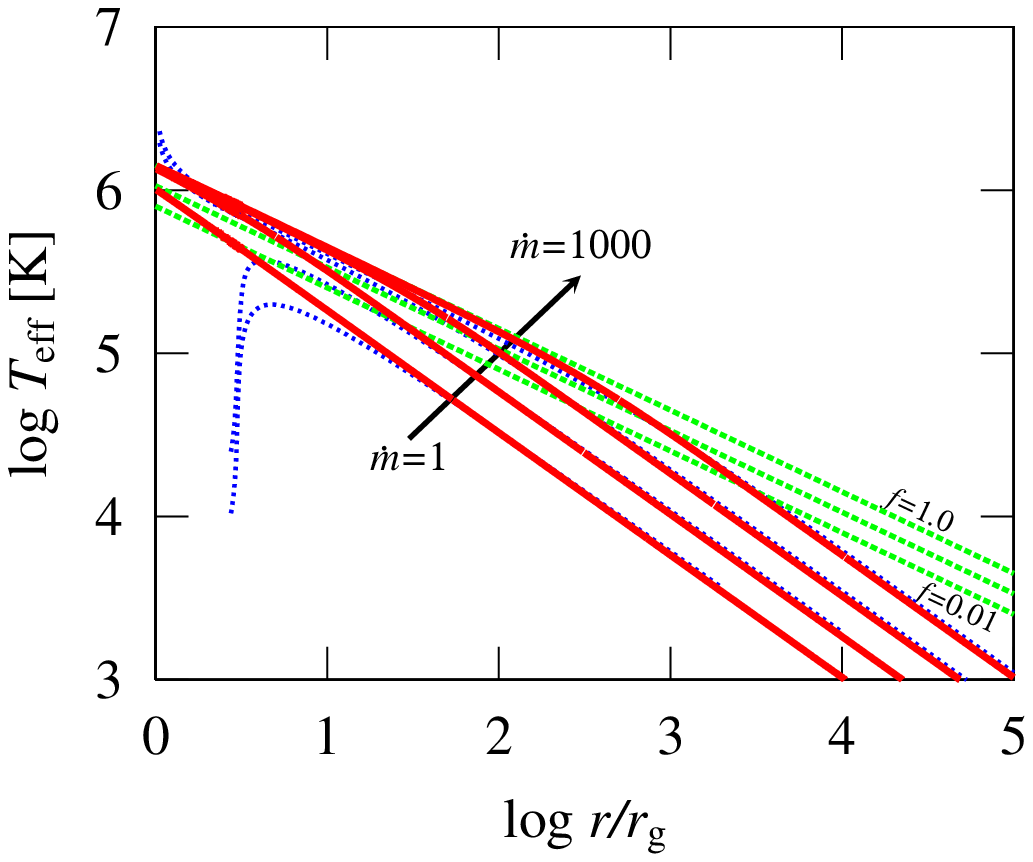}
  \end{center}
  \caption
{Same as Fig 2. but for the case where $m=10^6$. 
The  viscosity parameter is set to be $\alpha=0.1$. }
  \label{fig:smbh}
\end{figure}

As black hole mass increases, the size of the radiation-pressure-dominated region also increase. 
In figure \ref{fig:smbh}, we demonstrate the supermassive black hole case.
According to KFM98, the boundary radius, $\hat{r}_{\rm ab}$,
 between the radiation-pressure dominated region and 
 the gas-pressure dominated region is given by equation (\ref{rab}); 
$\hat{r}_{\rm ab} \sim 1800 (\alpha/0.1)^{2/21} (m/10^6)^{2/21} (\dot{m}/100)^{16/21}$
 for a supermassive black hole, i.e., an increase in black hole mass corresponds
 to an extension of the radiation-dominated region. 
Figure \ref{fig:smbh} shows that the analytic solutions for supermassive black holes are a better approximation than those for stellar mass black holes.  
However, it is necessary to consider carefully the effects of gravitational instability
 and self-gravity in supercritical accretion flows for supermassive black holes. 
We discuss these issues in section \ref{grav}.

\section{Discussion}
\label{sec:discussion}

\subsection{Gravitational Instability: Toomre's Q Parameters}
\label{grav}
Gravitational instability in accretion disks is an important issue
 in understanding the mass supply mechanism from the outer regions of the disk. 
A great amount of study has been conducted in this area 
 (e.g., Paczy\'{n}ski 1978; Lin \& Pringle 1987; Hur\'{e} et al. 1998; Kawaguchi et al. 2004).
In particular, Kawaguchi et al. (2004) investigated
 the effect of self-gravity in supercritical accretion flows 
 and concluded that the origin of the thermal component of a NLS1
 observed in infrared band is the outer self-gravitating disk. 
In general, the gas density in a supercritical accretion disk increases so that 
 the self-gravity effect may become important. 
The gravitational instability also depends on the value of $\alpha$, 
 i.e., if $\alpha$ is small ($\alpha \sim 0.01-0.001$), the disk mass increases. 

To examine the gravitational stability of an accretion flows, 
 we introduce Toomre's $Q$ parameter (Toomre 1964), 
\begin{equation}
 Q = \frac{c_{\rm s} \Omega}{\pi G \Sigma} \simeq 3.77 \times 10^{21} \alpha m^{-1} \dot{m}^{-1} f^{3/2} \hat{r}^{-3/2} 
\end{equation}
where $G$ is the gravitational constant. 
When $Q$ is less than unity, the disk is gravitationally unstable. 
We can consider the $Q=1$ radius; 
\begin{equation}
\label{q1}
 \hat{r}_{Q=1} \approx 5.21 \times 10^{5} \left( \frac{\alpha}{0.1} \right)^{2/3} \left( \frac{m}{10^7} \right)^{-2/3} \left( \frac{\dot{m}}{100} \right)^{-2/3} \left( \frac{f}{0.01} \right).   
\end{equation}
This radius is much larger than the gas-pressure-dominant radius $\hat{r}_{\rm ab}$, 
 which means that the disk is stable to gravitational perturbations
 in the radial direction. 
The $Q$ parameters for $10^6 M_\odot$ are plotted in figure 6. 
The $Q$ values for our new solutions are all larger than unity over a wide range in disk radius.  
Therefore, gravitational instability is not present the inner or middle disk regions for the range of disk parameters discussed here. 
Our result is supported by Kawaguchi et al. (2004) who also calculated $Q$ values for NLS1s,
 and found that $Q$ is greater than unity for all disk radii. 
Note that for more massive black holes ($M > 10^6 M_\odot$) or relatively low viscosity parameters ($\alpha < 0.1$),
 the $Q=1$ radius moves inward due to increasing gas density.
 (The $Q$ value for stellar mass black holes is much larger than unity.)

\begin{figure}
  \begin{center}
     \epsscale{1.0}
    \plotone{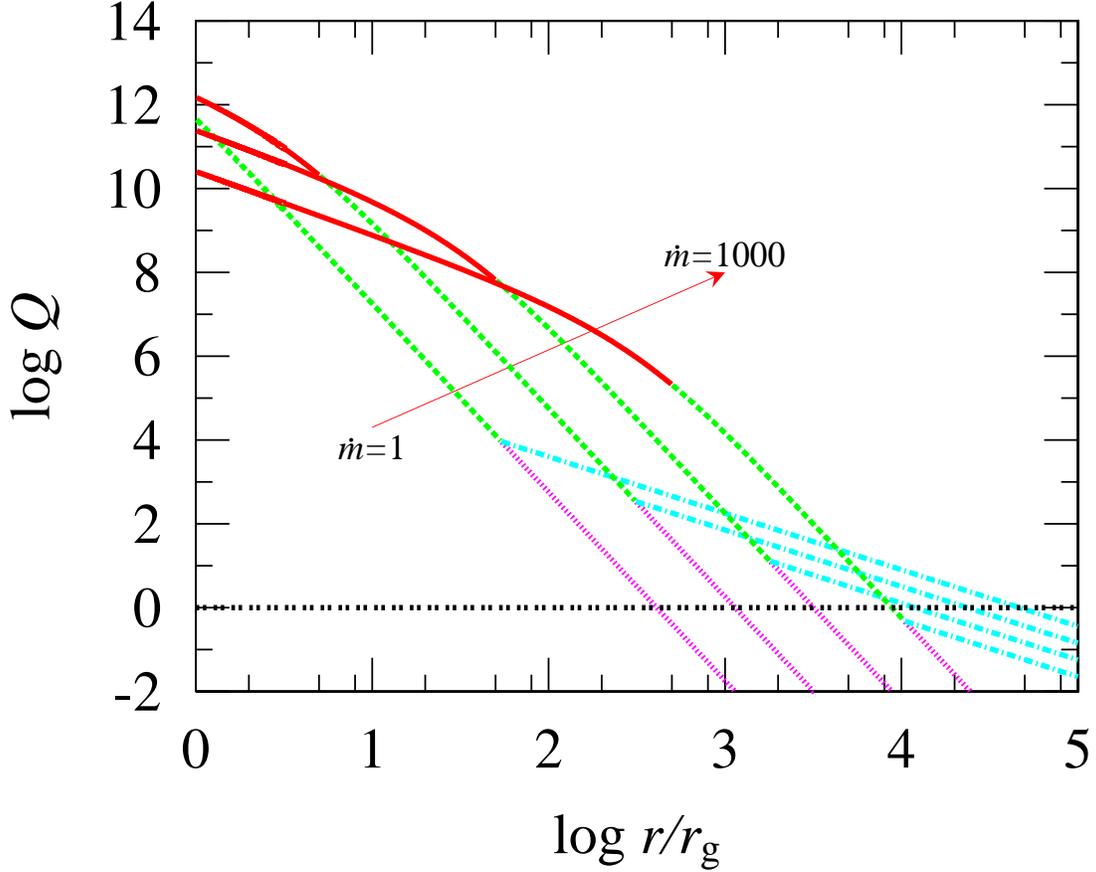}
  \end{center}
  \caption
{Toomre's $Q$ for different accretion rates
 ($\dot{m}$=1, 10, 100, and 1000 from bottom to top). 
In all cases, the black hole mass and the accretion rate are $m=10^6$ and $\alpha=0.1$. 
Solid lines, dashed lines show the advection-dominated regime and the radiation-dominated regime, respectively. 
Dot-dashed lines represent the gas-pressure dominated solutions. 
}
  \label{fig:qprm}
\end{figure}

\subsection{Total Luminosity}

Assuming spherical accretion,
 the maximum luminosity is determined
 by the balance between the gravitational force of gas and the radiation force. 
The mass accretion stops at the maximum luminosity, 
and this is the definition of the Eddington luminosity. 
In contrast, for disk accretion the gravitational force
 does not balance to the radiation force. 
In disk accretion, radiation escapes toward the vertical direction of the disk,
 so that the gravitational force cannot balance the radiation force. 
Eventually the mass accretion process leads to the production of large numbers of photons. 
The pressure gradient force ($\frac{1}{\rho} \frac{dp}{dr}$) is proportional to $r^{-2}$,
 which is the same radial dependence as gravity. 
The ratio of the gravity to the pressure gradient force is
 therefore always larger than unity,
 and hence the super-Eddington luminosities are possible. 
Numerical experiments indicate a maximum luminosity of $\sim 10 L_{\rm E}$
 for $\dot{m} = 10^3 -10^4$ (Abramowicz et al. 1988; Watarai et al. 2000).

We confirmed that our solutions are valid over a wide range of radii (see figure 7). 
In particular, the effective temperature from our analytic solution
 is in good agreement with numerical solutions. 
Accordingly, we can safely estimate the bolometric luminosity $L$ at high accretion rates
 using our analytic solutions. 
A total luminosity fitting formula for supercritical accretion
 was first derived by Watarai et al. (2000). 
Fukue (2004) has expressed the analytical form of $L$ using the self-similar formalism. 
Below we review again the self-similar determination of disk luminosity. 

\begin{figure}
  \begin{center}
    \epsscale{1.0}
    \plotone{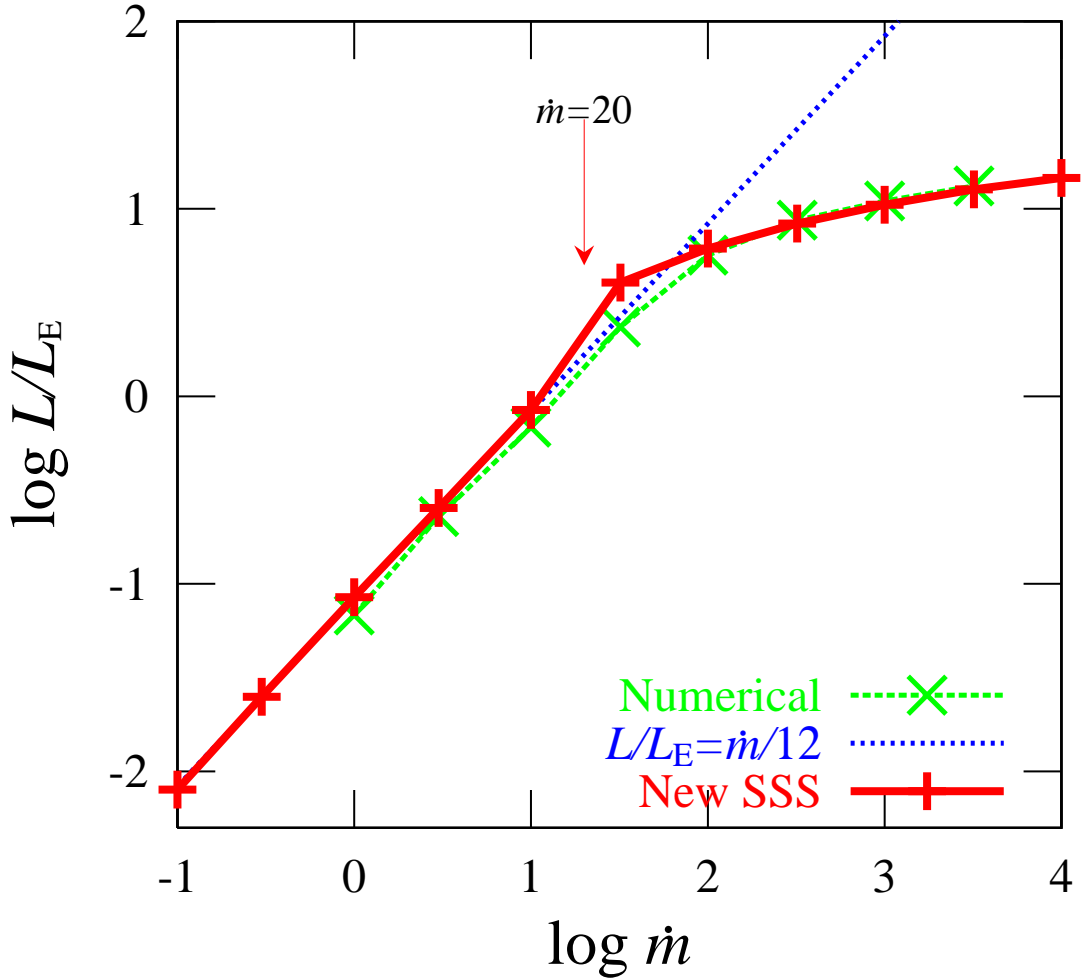}
  \end{center}
  \caption{Bolometric luminosity as a function of mass accretion rate. 
The solid line represents the total luminosity using equation
 (\ref{lbol3}) for $\dot{m} \geq 20$,
 and equation (\ref{lbol2}) for small accretion rates ($\dot{m} < 20$). 
The dashed and dotted lines are the numerical results and standard disk luminosity, 
 $L/L_{\rm E}=\frac{\dot{m}}{12}$, respectively. 
}
  \label{fig:lmdot}
\end{figure}

The luminosity is mainly produced by two components; 
 the inner slim disk $L_{\rm SLIM}$ and the outer standard disk $L_{\rm SSD}$: 
\begin{eqnarray}
L &=& L_{\rm SLIM} + L_{\rm SSD} \nonumber \\
  &=& \int_{r_{\rm in}}^{r_{\rm trap}} 2F_{\rm slim} 2\pi r dr 
    + \int_{r_{\rm trap}}^{r_{\rm out}} 2F_{\rm SSD} 2\pi r dr \nonumber \\
  &=& L_{\rm E} \frac{4}{I_4}\sqrt{\frac{B \Gamma_{\Omega}}{(2N+3)\xi \Omega_0^2}}
  f^{1/2} \ln\left[\frac{\rtrap}{r_{\rm in}}\right]   
  + \frac{GM\dot{M}}{2 \rtrap} 
\label{lbol1}
\end{eqnarray}
The bolometric luminosity normalized by the Eddington luminosity is 
\begin{equation}
L/L_{\rm E}  \approx
\left\{ \begin{array}{lcl}
4.0 f^{1/2} \ln\left[\frac{r_{\rm trap}}{r_{\rm in}}\right] + \frac{\dot{m}}{4 \hat{r}_{\rm trap}} & \mbox{$\dot{m} \gtrsim 20$} & \\
\frac{\dot{m}}{4 \hat{r}_{\rm trap}} & \mbox{$\dot{m} \lesssim 20$} & 
\end{array} \right.
\label{lbol2}
\end{equation}
We divided the calculation into two regimes in order to avoid $f < 0$. 
The disk's inner edge was set to 3.0 $\rg$ for simplicity. 
The previous luminosity estimate derived by Wang and Zhou (1999) 
never exceeds the Eddington luminosity. 
The discrepancy between our results and theirs is due to 
 a numerical coefficient and the second term of equation (\ref{lbol1}), which
 is the contribution of a standard disk. 
The bolometric luminosity for various accretion rates is shown in figure \ref{fig:lmdot}. 

In addition, we can calculate the bolometric luminosity using the new analytic solutions. 
The value of $f$ is a function of the radius and the mass accretion rate,
 and so the $f(\hat{r}, \dot{m})$ should be included in the flux integration;  
\begin{eqnarray}
L/L_{\rm E} &=&
  \frac{2}{I_{4}} \sqrt{\frac{B \Gamma_{\Omega}}{(2N+3)\xi \Omega_0^2}} 
  \int_{\hat{r}_{\rm in}}^{\hat{r}_{\rm out}} f(\hat{r}, \dot{m})^{1/2} \hat{r}^{-1} d\hat{r} \\
  & \approx & 2.01 \int_{\hat{r}_{\rm in}}^{\hat{r}_{\rm out}} f(\hat{r}, \dot{m})^{1/2} \hat{r}^{-1} d\hat{r}. 
\label{lbol3}
\end{eqnarray} 
However, if we use equation (\ref{lbol3}) for small accretion rates
 ($\dot{m} \lesssim 20$), the derived luminosities are significantly higher than the numerical results. 
This is because of the effect of the boundary condition for the inner region. 
As seen in figure \ref{fig:profile1}, 
 the difference between the numerical and analytical solutions increases for small $\dot{m}$. 
Overall, equation (\ref{lbol2}) gives a more accurate luminosity 
 compared to equation (\ref{lbol3}). 
If we consider the boundary condition more carefully, 
 this discrepancy can be reduced. This will form part of future work. 

We note that the observed luminosity depends on the viewing angle. 
In the disk geometry, the emitting direction is mainly toward the vertical direction
 of the accretion disk.   
Thus, a super-Eddington luminosity will only be observed when viewing the disk face-on,
 whereas it will be measured to be sub-Eddington if viewed edge-on. 
It may be that the bolometric luminosity locally exceeds the Eddington luminosity in 1D model. 
We should therefore take into account the effects of mass-loss to retain the self-consistency. 

The photon bubble mechanism is another promising model to produce super-Eddington luminosity (Arons 1992; Gammie 1998; Begelman 2002). 
The photon bubble instability leads to enhanced vertical transport of energy,
 so that the disk is efficiently cooled/or stabilized by the decreasing of thermal energy (Gammie 1998). 
However, numerical modeling of the photon bubble instability for high mass accretion rates has not yet been carried out
 (see Turner et al. 2005 for subcritical rates). 
It is unclear whether the photon bubble mechanism can be effective in regions with large amounts of gas and high radiation densities.  
The effect of photon bubble processes in supercritical accretion flows is an important area of future investigation.

\subsection{Stability of Supercritical Accretion Flows}

The standard disk is always secularly/thermally unstable in the radiation-pressure-dominant regime
 in classical studies (i.e., Lightman \& Eardley 1974; Shibazaki \& H\={o}shi 1975; Shakura \& Sunyaev 1976). 
However, the stability analysis of this early work is only applicable to local stability;
 the early work assumed that the viscous heating energy is locally balanced by radiative cooling,
 and neglected the effect of advective energy transport, which is important in high mass accretion rate regimes. 
Therefore, these classical treatments donot simply apply to the present study. 
Global disk stability at high mass accretion rates has been studied by several authors. 
Linear stability analysis for high mass accretion flows indicates that the secular/thermal modes
 do not grow over a wide range in radius (Honma et al. 1991; Wallinder 1991a, 1991b). 
Moreover, the acoustic mode is also stabilized by advective motion (Fujimoto \& Arai 1998). 
Honma et al. (1991) also performed 1D numerical simulation for relatively
 high mass accretion rates ($\dot{M}=10\dot{M}_{\rm Edd}$,
 where $\dot{M}_{\rm Edd}$ is the Eddington accretion rate,
 $\dot{M}_{\rm Edd}=16 L_{\rm E}/c^2$ for pseudo-Newtonian potential),
 and confirmed that the middle (unstable) region of the disk is stabilized
 by the inner advection-dominated regime and outer gas-pressure dominated regime. 
Therefore, our new solutions are globally stable against secular/thermal instabilities. 
Recent 2D radiation hydrodynamic (RHD) simulations by Ohsuga et al. (2005) showed that
 a quasi-steady disk structure appears for high accretion rates (e.g., $\dot{m}$= 100). 
Two-dimensional effects, e.g., convective motion, or outflows,
 may play an important role in preventing thermal instability. 
The photon bubble mechanism also stabilizes an unstable disk
 because it increase the radiative flux (Gammie 1998; Begelman 2002). 
Numerical experiments by Turner et al. (2006) have confirmed that
 the photon bubble mechanism produces an inhomogeneous density contrast,
 so that the disk can eventually radiate a large quantity of high energy photons
 from the disk equatorial plane effectively. 
 It is difficult to full consider such a detailed scenario at this stage. 
In-depth comparisons with 1D and 2D simulation results are required to confirm the stabilization
 mechanism of the thermal/secular instability. 
This will be the subject of future work.

\subsection{Is the Analytical Formula Useful for Observations?}

The greatest advantage of the analytical solutions presented here lies
 in their simplicity. 
However, analytical solutions may not be sufficiently accurate for detailed 
comparisons with observations because they do not incorporate
the effects of boundary conditions. 
Nevertheless, we have seen that the global properties of the flow dynamics
 and spectra obtained from our solutions are roughly consistent with those obtained using numerical
solutions.  
It is therefore worthwhile to discuss their observational implications. 

For radii less than the trapping radius,
 the temperature profile follows $T_{\rm eff} \propto r^{-0.5}$. 
and the spectral slope in this regime will follows $\nu F_{\nu} \propto \nu^0$. 
For radii larger than the trapping radius, 
 the temperature profile of a standard disk, 
 $T_{\rm eff} \propto r^{-0.75}$, is expected,
 and the resultant spectral slope will be $\nu F_{\nu} \propto \nu^{4/3}$. 
A spectrum observed to be flatter than the standard model would
 therefore be a useful tool for estimating the mass accretion rate. 

First, equation (\ref{trapr}) shows that the trapping radius can be determined
 from the normalized accretion rate $\dot{m}$. 
The temperature profile for high $\dot{m}$ disks changes at $r = r_{\rm trap}$; 
 therefore, the spectral slope must change at a characteristic frequency. 
If we can observe the turnover frequency in an accretion disk spectrum,
 we can estimate the accretion rate directly.
The normalized photon-trapping radius is defined in this paper as
 $\hat{r}_{\rm trap} \sim \dot{m}/2$,
 and the effective temperature at $\hat{r}_{\rm trap}$ is written as 
 a function of $f$, $m$, and $\dot{m}$.   
\begin{equation}
T_{\rm eff} (r_{\rm trap}) \approx  
\left\{ \begin{array}{lcl}
 1.33 \times 10^6 K
 \left(\frac{f}{0.5}\right)^{1/8}
 \left(\frac{m}{10}\right)^{-1/4}
 \left(\frac{\dot{m}}{100}\right)^{-1/2}  
 & \mbox{for stellar-mass BH} & \\
 7.50 \times 10^4 K
 \left(\frac{f}{0.5}\right)^{1/8}
 \left(\frac{m}{10^6}\right)^{-1/4}
 \left(\frac{\dot{m}}{100}\right)^{-1/2}  
 & \mbox{for super-massive BH} & 
\end{array} \right.
\end{equation}
By adapting Wien's displacement law ($\nu_{\rm max}=5.88 \times 10^{10} T$), 
 and after considering the spectral hardening factor $\kappa$ (Shimura \& Takahara 1995),
 we find the turnover frequency $\nu_{\rm to}$ to be  
\begin{equation}
\nu_{\rm to} \equiv \nu_{\rm max} \approx
\left\{ \begin{array}{lcl}
 1.33 \times 10^{17} {\rm Hz}
 \left(\frac{f}{0.5}\right)^{1/8}
 \left(\frac{m}{10}\right)^{-1/4}
 \left(\frac{\dot{m}}{100}\right)^{-1/2}  
 \left(\frac{\kappa}{1.7}\right)  
 & \mbox{for stellar-mass BH} &  \\
 7.49 \times 10^{15} {\rm Hz}
 \left(\frac{f}{0.5}\right)^{1/8}
 \left(\frac{m}{10^6}\right)^{-1/4}
 \left(\frac{\dot{m}}{100}\right)^{-1/2}  
 \left(\frac{\kappa}{1.7}\right)  
 & \mbox{for super-massive BH} & 
\end{array} \right.
\label{eq:nuto}
\end{equation}

For stellar mass black holes, the turnover frequency is located in
 the soft X-ray band ($\sim 0.54$ keV). 
As the mass accretion rate increases, 
 $\nu_{\rm to}$ shifts to lower energies,
 so that it may be marginally possible to measure after considering soft X-ray absorption. 
According to previous work, the spectral hardening factor, $\kappa$, may be larger than $1.7$ 
 in supercritical accretion flows (Kawaguchi 2003; Shimura \& Manmoto 2003),
 in which case it will be possible to estimate the accretion rate robustly; 
 however, it will also be necessary to carefully consider 
 the effects of Comptonization.  

In supermassive black holes ($M \gtrsim 10^6 M_\odot$) 
 it may be possible to measure the turnover frequency. 
The position of the peak blackbody emission is located at $\sim 10^{14-15}$ Hz;
 which is within the soft UV/optical bands. 
Even if the black hole mass or accretion rate increases further, 
 the turnover frequency $\nu_{\rm to}$ still appears in the IR band; 
 however, the big blue bump which is observed in the near-infrared band in quasar spectra
 cannot usually be observed directly due to the presence of strong hot dust emission
 (Chiang \& Blaes 2003; Kishimoto et al. 2005). 
From a theoretical point of view, the effects of Comptonization, metal abundance,
 non-LTE, and a hot coronal component can all strongly affect to the big blue bump component
 in the quasar/AGN spectra (Hubeny et al. 2001). 
The determination of accretion disk parameters using the photon trapping radius
 may therefore be extremely difficult in practice. 

The origin of the high energy Wien spectrum is 
 unknown due to the complexity of physical processes, 
for example, the relativistic Doppler effect, 
 gravitational redshift, photon trapping, and Comptonization, which
 are expected to couple together. 
However, measurements of the turnover frequency could be a useful method to search
 for supercritical accretion candidates,
 or to obtain the information about the mass accretion rate.
Using equation (\ref{eq:nuto}), we can estimate the mass accretion rate, 
\begin{equation}
 \dot{m} = 1.95 \times 10^3 \left(\frac{f}{0.5}\right)^{1/4}
\left(\frac{m}{10^6}\right)^{-1/2}
\left(\frac{\nu_{\rm to}}{10^{15} {\rm Hz}} \right)^{-2}. 
\end{equation}
Unfortunately, the use of the photon-trapping radius
 to determine the disk parameters may be extremely difficult
 due to both theoretical and observational difficulties.

\section{Conclusions}

\noindent{1.}
We determined new analytic solutions to describe high luminosity accretion flows. 
Our analytic solutions are roughly consistent with the global
numerical solutions for the same assumptions (except at the inner or outer boundary). 
In particular, we explicitly showed that
 the ratio of the advective cooling rate to the viscous heating rate
 $f$ is a function of the radius and mass accretion rate. 

\noindent{2.}
 Thermodynamic related quantities obtained using the analytic solutions,
 in particular, the temperature distribution gives a reasonable approximation
 to the profile of the whole disk.  
This demonstrates that our solutions are useful for 
interpreting observations, although complications may arise due to
emission and scattering in the disk corona and possible Compton scattering.  

\noindent{3.}
Using observations of the flux ratios of particular objects, we can discriminate 
between a standard disk and a slim disk. 
In other words, this is an indirect method of measuring the accretion rate.
For this measurement, the key physical quantity is again the photon-trapping radius.
Hence, the estimation of the trapping radius is of greatest 
theoretical importance. Our model is a simple 1D model, 
 and so it necessary to calculate the exact trapping radius using more than
 2D or 3D radiation hydrodynamics. 

In the present study, we did not consider the effect of outflows. 
In high mass accreting flows, the gravitational force in the vertical direction of the disk cannot balance
 the flux force, and in some cases, the flux locally exceeds the Eddington limit. 
Hence, the vertical structure of supercritical accretion flows
 will also be an important topic for future studies. 

Two-dimensional RHD simulations are technically complex and Time consuming tasks.
Our new analytic solutions for optically thick disks will therefore be useful
 in developing a new understanding of luminous black hole candidates. 
We expect to use the future observational data
 to investigate the supercritical accretion scenario in more detail. 

\acknowledgments
We would like to thank R. Narayan, S. Mineshige, J. Fukue, H. Kamaya,
 K. Ohsuga, and A.K. Inoue for helpful comments and discussions.
The author also would like to thank C. James for checking the manuscript,
 and thanks the Yukawa Institute for Theoretical Physics at Kyoto
University, where this work was initiated during the YITP-W-01-17 on
``Black Holes, Gravitational Lens, and Gamma-Ray Bursts''. 
This work was supported in part by the Grants-in Aid of the
Ministry of Education, Science, Sports, and Culture of Japan
(16004706, KW).



\begin{thebibliography}{}
\bibitem[Abramowicz(1978)]{abra78} Abramowicz, M.A., Jaroszy\'{n}ski, M., Sikora, M. 1978, \aap, 63, 221
\bibitem[Abramowicz(1988)]{abra88} Abramowicz, M.A., Czerny, B., 
    Lasota, J.P., Szuszkiewicz, E. 1988, \apj, 332, 646
\bibitem[Wallindar (1991a)]{wallin91a} Wallinder, F.H. 1991a, \aap, 249, 107
\bibitem[Wallindar (1991b)]{wallin91b} Wallinder, F.H. 1991b, \aap, 253, 184
\bibitem[Arons (1992)]{arons92} Arons, J. 1992, \apj, 388, 561
\bibitem[]{}Ball, G.H., Narayan, R., Quataert, W. 2001, \apj, 552, 221 
\bibitem[]{}Begelman, M.C. 1978, \mnras, 184, 53 
\bibitem[]{}Begelman, M.C., Meyer, D.L. 1982, \apj, 253, 873 
\bibitem[Begelman (2002)]{begel02} Begelman, M.C. 2002, \apj, 568, L97 
\bibitem[]{}Bian W., Zhao, Y. 2004, \mnras, 352, 823 
\bibitem[]{}Blandford, R.D., Begelman, M.C. 1999, \mnras, 303, L1 
\bibitem[]{} Collin, S., Kawaguchi, T. 2004, \aap, 426, 797
\bibitem[]{} Chandrasekhar, S. 1967, {\it Stellar Structure} (The Univ. of Chicago Press, Chicago)
\bibitem[]{} Chiang, J., Blaes, O. 2003, \apj, 586, 97
\bibitem[]{ebi03} Ebisawa, K., \.{Z}ycki, P.T, Kubota, A., Mizuno, T., \& Watarai, K. 2003, \apj, 597, 780
\bibitem[Fujimoto (1998)]{fuji98} Fujimoto, S., Arai, K. 1998, \aap, 330, 1190
\bibitem[Fukue (2000)]{fuku00} Fukue, J. 2000, \pasj, 52, 829 
\bibitem[Fukue (2004)]{fuku04} Fukue, J. 2004, \pasj, 56, 569 
\bibitem[Gammie (1998)]{gam98} Gammie, C. 1998, \mnras, 297, 929 
\bibitem[Honma et al(1991a)]{honma91a} Honma, F., Matsumoto, R., Kato, S. 
    1991, \pasj, 43, 147 
\bibitem[Honma et al(1991b)]{honma91b} Honma, F., Matsumoto, 
    R., Kato, S., Abramowicz, M.A. 1991, \pasj, 43, 261 
\bibitem[Honma (1996)]{honma96} Honma, F. 1996, \pasj, 48, 77 
\bibitem[]{} H\={o}shi, R. 1977, $Prog. Theor. Phys.$, 58, 1191
\bibitem[]{} Hubeny, I., Blaes, O., Krolik, J.H., Agol, E. 2001, 559, 680
\bibitem[]{} Hur\'{e}, J.-M. 1998, \aap, 337, 625
\bibitem[]{jaro80} Jaroszy\'{n}ski, M., Abramowicz, M.A., Paczy\'{n}sky, B. 1980, Acta Astronomica, 30, 1
\bibitem[Kato(1998)]{kato98} Kato, S., Fukue, J., Mineshige, S. 
    1998, Black-Hole Accretion Disks (Kyoto University Press, Kyoto) (KFM98)
\bibitem[]{} Kawaguchi, T., Pierens, A., Hur\'{e}, J.-M. 2004, \aap, 415, 47
\bibitem[]{} Kawaguchi, T. 2003, \apj, 593, 69
\bibitem[]{} Kishimoto, M., Antonucci, R., Blaes, O. 2005, \mnras, 364, 640
\bibitem[]{} Lin, D.N.C, Pringle, J.E. 1987, \mnras, 225, 607
\bibitem[]{} Lightman, A.P, Eardley, D.M. 1974, \apj, 187, L1
\bibitem[Matsumoto(1984)]{matu84} Matsumoto, R., Kato, S., 
         Fukue, J., Okazaki, A. 1984, \pasj, 36, 71
\bibitem[Mineshige et al.(2000)]{mine00} Mineshige, S., Kawaguchi, T., 
         Takeuchi, M., Hayashida, K. 2000, \pasj, 52, 499
\bibitem[]{} Narayan, R., Yi, I. 1994, \apjl, 428, L13
\bibitem[]{} Narayan, R., Yi, I. 1995, \apj 452, 710
\bibitem[]{} Narayan, R., Kato, S., Honma, F. 1997, \apj, 476, 49
\bibitem[]{} Narayan, R., Igumenshchev, I.V., Abramowicz, M.A. 2000, \apj, 539, 798
\bibitem[Ohsuga et al.(2005)]{oh05} Ohsuga, K., Mori, M., Nakamoto, T., 
         Mineshige, S. 2005, \apj, 628, 368 
\bibitem[Paczynski (1980)]{paczyn80} Paczy\'{n}sky, B., Wiita, P.J. 
         1980, \aap, 88, 23 
\bibitem[]{} Quataert, E., Gruzinov, A. 
         2000, \apj, 539, 809 
\bibitem[Shakura \& Sunyaev.(1973)]{ss73} Shakura, N.I., Sunyaev, R.A. 
         1973, \aap, 24, 337
\bibitem[Shakura \& Sunyaev.(1976)]{ss76} Shakura, N.I., Sunyaev, R.A. 
         1976, \mnras, 175, 613
\bibitem[]{} Shibazaki, T., H\={o}shi, R. 1975, Prog. Theor. Phys., 54, 706
\bibitem[Shimura and Takahara 1995]{st95} Shimura, T., \& Takahara, F. 
    1995, \apj, 445, 780
\bibitem[Shimura and Manmoto 2003]{sm03} Shimura, T., \& Manmoto, T.
    2003, \mnras, 338, 1013 
\bibitem[]{soria} Soria, R., Kong, A.K.H. 2002, \apjl, 572, L33
\bibitem[Spruit et al.(1989)]{sp1989} Spruit, H.C., Matsuda, T., Inoue,
	 M., Sawada, K. 1989, \mnras, 229, 517 
\bibitem[Szuszkiewicz et al.(1996)]{szu96} Szuszkiewicz, E., 
         Malkan, M.A., Abramowicz, M.A. 1996, \apj, 458, 474
\bibitem[Toomre (1964)]{toom64} Toomre, A. 1964, \apj, 139, 1217
\bibitem{Turner et al. (2005)}{turn05} Turner, N.J., Blaes, O.M., Socrates,
        A., Begelman, M.C., Davis, S.W. 2005, \apj, 624, 267
\bibitem[]{} Wang, J.M., Zhou, Y.Y. 1999, \apj, 516, 420
\bibitem[]{} Wang, J.M., Szuszkiewicz, E., Lu, F.J., Zhou, Y.Y., 1999, 
    \apj, 522, 839
\bibitem[Watarai(1999)]{wata99} Watarai, K., Fukue, J. 1999, \pasj, 51, 725
\bibitem[Watarai(2000)]{wata00} Watarai, K., Fukue, J.,
    Takeuchi, M., Mineshige, S. 2000, \pasj, 52, 133
\bibitem[]{} Watarai, K., Mizuno, T., Mineshige, S. 2001, \apjl, 549, 77
\bibitem[]{} Watarai, K., Mineshige, S. 2001, \pasj, 53, 915 
\end{thebibliography}
\end{document}